\documentclass[aps,prb,amsmath,amssymb,superscriptaddress,twocolumn]{revtex4}
\usepackage[english]{babel}
\usepackage{graphicx,bm,amssymb,amsmath}
\usepackage{color,wrapfig,braket}
\usepackage{hyperref,placeins,ulem}
\usepackage[caption=false]{subfig}
\usepackage{subfig}
\usepackage{ulem}

\usepackage{dsfont}
\usepackage{epsfig}
\usepackage{verbatim}
\usepackage{yhmath}
\usepackage{bbm}

\def\T{\intercal}
\def\SO{\textrm{SO}}
\def\O{\textrm{O}}
\def\R{\mathbb{R}}

\def\sgn{{\rm sgn}}

\def\J{\mathcal{J}}
\def\tO{\tilde{O}}

\makeatletter
\newcommand\xleftrightarrow[2][]{%
  \ext@arrow 9999{\longleftrightarrowfill@}{#1}{#2}}
\newcommand\longleftrightarrowfill@{%
  \arrowfill@\leftarrow\relbar\rightarrow}
\makeatother

\begin{document}

\title{Lattice Construction of Duality with Non-Abelian Gauge Fields in 2+1D }
\author{Chao-Ming Jian}
\affiliation{Station Q, Microsoft Research, Santa Barbara, California 93106-6105, USA}
\affiliation{Kavli Institute of Theoretical Physics, University of California, Santa Barbara, California 93106, USA}
\author{Zhen Bi}
\affiliation{Department of Physics, Massachusetts Institute of Technology, Cambridge, Massachusetts 02139, USA}
\author{Yi-Zhuang You}
\affiliation{Department of Physics, University of California, San Diego, CA 92093, USA}
\affiliation{Department of Physics, Harvard University, Cambridge, MA 02138, USA}

\date{\today}

\begin{abstract}
The lattice construction of Euclidean path integrals has been a
successful approach of deriving 2+1D field theory dualities with a
U$(1)$ gauge field. In this work, we generalize this lattice
construction to dualities with non-Abelian gauge fields. We
construct the Euclidean spacetime lattice path integral for a
theory with strongly-interacting $\SO(3)$ vector bosons and
Majorana fermions coupled to an $\SO(3)$ gauge field and derive an
exact duality between this theory and the theory of a free
Majorana fermion on the spacetime lattice. We argue that this
lattice duality implies the desired infrared duality between the
field theory with an $\SO(3)$ vector critical boson coupled to an
$\SO(3)_1$ Chern-Simons gauge theory, and a free massless Majorana fermion
in 2+1D. We also generalize the lattice construction of dualities to models with $\O(3)$ gauge fields.
\end{abstract}

\maketitle

\section{Introduction}

The classic particle-vortex duality~\cite{Peskin1978,Dasgupta1981}
of bosonic systems revealed that different Lagrangians can
secretly correspond to the same conformal field theory in the
infrared limit. Recently the boson-fermion
dualities~\cite{Polyakov1988,wufisher,Giombi2012,Aharony2012_1,Aharony2012_2,seiberg1},
fermion-fermion
dualities~\cite{Son2015,MetlitskiDualDirac,WangDualDirac,mrossduality},
and many descendant 2+1D field theory
dualities~\cite{WangO5,SSWW2016,Hsin2016,YouXu2015,Hsin2017,Aharony2016,Karch2016,Metlitski2017,Gaiotto2018,Gomis2017,KarchRobinson2017,GurAri2015,Benini2018,Jensen2017,Jensen2018,Cordova2017_1,Cordova2017_2,mengxu}
have attracted a lot of attentions. The importance of these
dualities are not just limited to their own theoretical interests.
They also have close connections to gauge-gravity duality
\cite{Giombi2012,Aharony2012_1,Aharony2012_2}, mirror symmetry in
supersymmetric field
theory~\cite{Kachru2017_1,Kachru2017_2,Karch2018}, fractional
quantum Hall
systems\cite{Fradkin1996,Son2015,Radicevic2016,Hui2018,WangSenthil2016,Sodemann2017}
as well as deconfined quantum critical
points~\cite{WangO5,dualnumerical}. These dualities have
tremendously advanced our understandings of 2+1D conformal field
theories. In fact, a large class of them are weaved into a duality
web through SL$(2,\mathbb{Z})$ transformations on U$(1)$ currents
and on the Chern-Simons terms of the U$(1)$ gauge
fields\cite{SSWW2016,Witten2003}. While the duality web itself
provides interesting cross-check among different dualities with
U$(1)$ gauge fields, analytical derivation for certain members of
the web are also obtained using wire constructions
\cite{Mross2016,Mross2017}, deformation of exact supersymmetric
dualities\cite{Kachru2017_2,Karch2018,GurAri2015} and lattice
constructions \cite{Dasgupta1981,Chen2018} of Euclidean path
integrals.


Although some of the descendant dualities mentioned above have
received rather positive numerical
evidences~\cite{dualnumerical,karthik2017}, rigorously speaking
most of the dualities are still conjectures. Apart from the
consistency check on global symmetries, anomalies and matching
global phase
diagrams~\cite{Hsin2016,Hsin2017,Aharony2016,Karch2016,Metlitski2017,Gaiotto2018,Gomis2017,KarchRobinson2017,GurAri2015,Benini2018,Jensen2017,Jensen2018,Cordova2017_1,Cordova2017_2},
rigorous analytical results are much harder to obtain for
dualities in the presence of non-Abelian gauge fields. Exact
results, say, on partition functions, operator scaling dimensions,
etc. in these dualities mostly rely on the large-$N$
limit\cite{Giombi2012,Aharony2012_1,Aharony2012_2,Banerjee2014,Jain2013,Maldacena2013}
or well-established level-rank dualities of Chern-Simons gauge
theories without dynamical matter fields
\cite{Naculich1990,Mlawer1991,Naculich2007}. Derivation of
dualities with non-Abelian gauge group of finite rank and with
dynamical matter fields is certainly of great importance.

Among all the approaches mentioned above, deriving the dualities
via the lattice construction of the Euclidean path
integral\cite{Dasgupta1981,Chen2018} stands out for its exactness
and explicitness in connecting the degrees of freedom on the two
sides of a duality without using any large-$N$ limit or
supersymmetry. This method often only relies on the mild
assumption that the theory defined on the lattice indeed flows
under renormalization group to its natural continuum limit in the
infrared (IR). Previously, the lattice construction has only been
carried out for dualities with U(1) gauge
fields\cite{Dasgupta1981,Chen2018}. In this paper, we generalize
the lattice construction to dualities with orthogonal nonabelian
gauge groups. We focus on the duality between a critical $\SO(3)$
vector boson coupled a $\SO(3)_1$ Chern-Simons gauge theory and a
free massless Majorana fermion in 2+1D, which was proposed in
Ref.~\onlinecite{Metlitski2017,Hsin2017}. We first construct the
Euclidean spacetime lattice path integral of a theory with
$\SO(3)$ vector bosons and Majorana fermions coupled to a $\SO(3)$
gauge field, which is connected to the theory of a single flavor
critical boson coupled to a $\SO(3)_1$ Chern-Simons term in the IR.
We perform an {\it exact} mapping of this spacetime lattice path
integral to that of a free Majorana fermion whose continuum limit
agrees with the expected dual of the critical bosons.
Generalizing this construction, we
obtain a slightly different lattice duality between an interacting
model with O(3) vector bosons and Majorana fermions coupled to
O(3) gauge fields and the theory of free Majorana fermions.

\section{Lattice Duality with \SO(3) gauge fields}
\label{Sec:SO(3)Duality} In Ref.
\onlinecite{Metlitski2017,Hsin2017}, a duality between the
$\SO(3)$ critical boson coupled to $\SO(3)_1$ Chern-Simons gauge
theory and the free gapless Marjorana fermions in 2+1D was
proposed. The continuum description of the boson side of the
duality is given by the Lagrangian:
\begin{align}
\mathcal{L}_b = & |(\partial_\mu-ia_\mu )\phi|^2  + r |\phi|^2 + g|\phi|^4
\nonumber
\\
& + \frac{i}{2\cdot 4\pi} \text{tr}_{\SO(3)} \left( a \wedge da -
\frac{2i}{3}a \wedge a \wedge a \right),
\label{Eq:BosonContinuumTheory}
\end{align}
where $\phi$ is a 3-component real vector field coupled to an
$\SO(3)$ gauge field $a$. 
The $\SO(3)$ gauge field $a$ is subject to a Chern-Simons term at
level-1 in which the trace ``$\text{tr}$" is taken in the vector
representation of $\SO(3)$. This theory has an dual description of
a single two-component Majorana fermion $\xi$ in 2 + 1D given by
the simple Lagrangian
\begin{align}
\mathcal{L}_f = & \bar{\xi} \gamma^\mu\partial_\mu\xi - m\bar{\xi}\xi.
\label{Eq:FermionContinuumTheory}
\end{align}
This duality is directly shown to hold in the gapped phases with
$r$ and $m$ carrying the same sign\cite{Metlitski2017,Hsin2017}.
It is further conjectured to be valid even at the critical point
where $r=m=0$. In this section, we will start with formulating a
lattice theory in close connection to the critical boson theory
and constructing its path integral on the Euclidean spacetime
lattice. We will then map this path integral exactly to that of a
free Majorana fermion. As we will see later this exact mapping is
valid regardless of whether the resulting phases are gapped or at
the critical point. We will also refer to this type of exact mapping between two lattice theories as a lattice duality. After establishing the lattice duality, we
will discuss the correlation functions and the $Z_2$ global
symmetry across the dualtiy.

\subsection{Euclidean Path Integral on the Lattice}
We start with the basic ingredients for constructing the spacetime
lattice path integral for the boson side of the duality. We
consider a discretized 3D Euclidean spacetime lattice, which is
taken to be a cubic lattice for simplicity. On each site $n$ of
the spacetime lattice, there is a 3-component unit vector $v_n =
(v_{n,1},v_{n,2},v_{n,3})^\T$ that represents the SO$(3)$ vector
boson fields. The vector boson fields couple to their nearest
neighbors and to the $\SO(3)$ gauge field residing on the links
that connect them, leading to the following contribution to the
Euclidean action:
\begin{align}
S_{\rm bg}[v,O] = \sum_n \sum_{\mu=x,y,z} -J~ v_{n+\mu,i} O_{ij}^{n\mu} v_{n,j},
\label{Eq: Action_SO(3)_vectors}
\end{align}
where $\mu=x,y,z$ is summed over the unit lattice vector along the
positive $x,y,$ and $z$ direction. $J$ is the coupling constant of
the SO$(3)$ vector bosons, which we assume to be always positive.
$O_{ij}^{n\mu}$ is an $\SO(3)$ matrix that represents the $\SO(3)$
gauge connection along the links between sites $n$ and $n+\mu$.
The repeated $\SO(3)$ vector/matrix indices $i,j$ are implicitly
assumed to be summed automatically from 1 to 3.

According to the Lagrangian Eq. \ref{Eq:BosonContinuumTheory}, we
also need to introduce a Chern-Simons term for the gauge field
$O^{n\mu}$. While it is difficult to directly write down the
Chern-Simons term on the lattice, we can circumvent the difficulty
by coupling the $\SO(3)$ gauge field to a massive Majorana
fermion. This method is a direct generalization of the
construction of the U(1) Chern-Simons term in a lattice duality
studied in Ref. \onlinecite{Chen2018}. In this method, the
Chern-Simons term can be viewed as the outcome of integrating out
the massive Majorana fermions. Following Wilson's
approach\cite{Wilson1974,Wilson}, we can write down the action for
lattice Majorana fermions coupled to the $\SO(3)$ gauge field:
\begin{align}
& S_{\rm fg}[\chi,O] =
\nonumber
\\
& \sum_n  \sum_{\mu = x,y,z } \bar{\chi}_{n+\mu,i} (\sigma^\mu -R) O^{n\mu}_{ij} \chi_{n,j}
+  M \sum_n \bar{\chi}_{n,i}  \chi_{n,i} ,
\label{Eq: Action_SO(3)_Majorana}
\end{align}
Again, the repeated $\SO(3)$ indices $i,j$ are automatically
summed over. For each site $n$ and each $\SO(3)$ color index $i$,
the fermion field $\chi_{n,i} = (\chi_{n,i,1}, \chi_{n,i,2})^{\T}$ is a 2-component spinor consists of two real Grassmann
numbers. $\sigma^\mu$ stands for the Pauli matrices.
$\bar{\chi}_{n+\mu,i}$ is defined as $\bar{\chi}_{n+\mu,i} =
\chi^{\T} \sigma^y$. When we turn off the gauge field
$O^{n\mu}$, the action Eq. \ref{Eq: Action_SO(3)_Majorana} alone
gives rise to $2^3=8$ Majorana fermions in the IR whose masses are
controlled by the parameters $R$ and $M$. The mass configuration
of the massive Majorana fermions determines the Chern number $C$
of the occupied bands of the Majorana fermions. To be more precise, by the band structure of Majorana fermions, we refer to the band structure obtained from quantizing the $8$ Majorana fermions. And by occupied bands, we mean all the negative energy states associated to the $8$ IR Majorana fermions after the quantization. The Chern number also serves as the level of the Chern-Simons term of the
$\SO(3)$ gauge field when the fermion field $\chi_n$ is integrated
out in the action $S_{\rm fg}[\chi,O]$. The relation between the
Chern number $C$ and parameters $R$ and $M$ that control the bare
band structure of the Majorana fermions field $\chi_n$ is given
by\cite{Golterman1993,Coste1989}
\begin{align}
C=\begin{cases}
2 \sgn(R) & 0 <|M| < |R|, \\
-\sgn(R) & |R|  <|M| < 3|R|, \\
0 & |M| > 3|R|. \\
\end{cases}
\label{Eq:Majorana_Chern_Number}
\end{align}
Naively, by combining the action Eq. \ref{Eq:
Action_SO(3)_vectors} and Eq. \ref{Eq: Action_SO(3)_Majorana}, one
would have already had all the essential ingredients for a lattice
version of the field theory Eq. \ref{Eq:BosonContinuumTheory}.
However, for reasons that will become clear later, we would also
like to include the interaction between the vector bosons and the
Majorana fermions:
\begin{align}
& S_{\rm int}[\chi, v] =  \frac{U_1}{4} \sum_{n,\mu}
\varepsilon_{ii'i''}\varepsilon_{jj'j''} \J^{n\mu}_{ij} \J^{n\mu}_{i'j'}
v_{n+\mu,i''} v_{n, j''}
\nonumber \\
& ~~~~ + \frac{U_2}{36}
\sum_{n,\mu}
\varepsilon_{ii'i''}\varepsilon_{jj'j''}
\J^{n\mu}_{ij} \J^{n\mu}_{i'j'} \J^{n\mu}_{i''j''} ,
\label{Eq:S_SO3_int_def}
\end{align}
where we've introduced the notation $\J^{n\mu}_{ij} \equiv
\bar{\chi}_{n+\mu,i} (\sigma^\mu -R) \chi_{n,j} $ for the fermion
current. $\varepsilon_{ii'i''}$ and $\varepsilon_{jj'j''}$
represent the totally-anti-symmetric tensor. $U_1$ and $U_2$ are
coupling constants of these interactions in $S_{\rm int}$. It is
straightforward to verify that $S_{\rm int}$ is invariant under
the $\SO(3)$ gauge transformations.

Now, we are ready to introduce our Euclidean spacetime lattice
path integral for one side of the duality:
\begin{align}
Z_b = \int D[O^{n\mu}] \int D[\chi_{n,i}] \int D[v_n] e^{-S_{\rm
bg} - S_{\rm fg} - S_{\rm int }},
\label{Eq:Lattice_Boson_Fermion_theory}
\end{align}
where the integration of the gauge field $\int D[O^{n\mu}]$ is
implemented as the integration of the matrix $O^{n\mu}$ on each
link over the Haar measure of $\SO(3)$. The boson field $v_n$ on
each site is integrated over the unit 2-sphere $S^2$, while the
fermion fields $\chi_{n,i}$ are integrated as real Grassmann
numbers. As we discussed, when the coupling constants $U_1$ and
$U_2$ are set to zero, after the Majorana fermions are integrated
out, this model naturally realizes the theory of $\SO(3)$ vector
bosons coupled to a $\SO(3)$ gauge field with the Chern-Simons
level given by Eq. \ref{Eq:Majorana_Chern_Number}. When $U_1$ and
$U_2$ are finite, the relation between the Chern-Simons level
generated by the Majorana fermions and parameters $M$ and $R$ is
expected to be modified. We will address the effect of $S_{\rm int}$
and how it alters the interpretation of the lattice path integral
$Z_b$ as we proceed in obtaining its dual theory.

\subsection{Exact Mapping of Lattice Path Integral}
\label{Sec:ExactMap}
We notice that the $\SO(3)$ gauge field $O^{n\mu}$ in Eq.
\ref{Eq:Lattice_Boson_Fermion_theory} can be integrated out
analytically. In order to do so, we only need to consider the
$S_{\rm bg}[v,O] $ and $S_{\rm fg}[\chi,O] $ parts of the action:
\begin{align}
& \int D[O^{n\mu}] ~ e^{-S_{\rm bg}[v, O]} e^{ - S_{\rm fg}[\chi, O] }
\nonumber \\
& = \exp \left( -\sum_n M \bar{\chi}_{n,i} \chi_{n,i}\right)
\times \prod_{n,\mu}  \left[ \sum_{l=0}^\infty \sum_{m=0}^\infty
\frac{J^l}{l!} \frac{(-1)^m}{m!}
 \right.
\nonumber \\
& ~~~~ \left. \int dO^{n\mu} \left(v_{n+\mu,i} O^{n\mu}_{ij}
v_{n,j} \right)^l \left( \bar{\chi}_{n+\mu,i} (\sigma^\mu -R)
O^{n\mu}_{ij} \chi_{n,j} \right)^m \right],
\end{align}
From now on, we will set $R=-1$ which renders the matrices
$\sigma^\mu-R$ rank 1 and which combined with the fermionic
statistics of the Grassmann numbers leads to vanishing
contributions for all the terms with $m>3$ (see Appendix \ref{App:IntegrationSO(3)} for a more detailed explanation). Notice that the
integration over all gauge field configurations $\int D[O^{n\mu}]$
factorizes into the integration of SO(3) matrices $O^{n\mu}$ under
the Haar measure on each link. After conducting a term by term
integration, we obtain that
\begin{align}
& \int D[O^{n\mu}] ~ e^{-S_{\rm bg}} e^{ - S_{\rm fg}} =
\left(\frac{\sinh J}{J}\right)^{3N_s} e^{-S_{\rm bfg}[v,\chi]}
\end{align}
where $N_s$ is the number of sites in the Euclidean spacetime lattice
and the effective action $S_{\rm bfg}[v,\chi]$ takes the form:
\begin{align}
S_{\rm bfg}[v,\chi] = & \sum_{n}  M \bar{\chi}_{n,i} \chi_{n,i}
\nonumber
\\
&+\sum_{n,\mu} \Big\{ K (\bar{\chi}_{n+\mu,i} v_{n+\mu,i})
(\sigma^\mu -R) (\chi_{n,j} v_{n,j})
\nonumber \\
&- \frac{K}{4} \varepsilon_{ii'i''}\varepsilon_{jj'j''}
\J^{n\mu}_{ij} \J^{n\mu}_{i'j'} v_{n+\mu,i''} v_{n, j''}
\nonumber \\
& \left.
+
 \frac{1-K^2}{36}
\varepsilon_{ii'i''}\varepsilon_{jj'j''}
\J^{n\mu}_{ij} \J^{n\mu}_{i'j'} \J^{n\mu}_{i''j''}
\right\}.
\label{Eq:Eff_Action_From_SO3}
\end{align}
with $K = \frac{J \cosh J - \sinh J}{J \sinh J}$ which is a
positive number between 0 and 1 for all $J>0$. The details of the derivation of Eq. \ref{Eq:Eff_Action_From_SO3} can be found in Appendix \ref{App:IntegrationSO(3)}. Notice that the
effective action $S_{\rm bfg}[v,\chi]$ contains similar
interactions to the ones we introduce in $S_{\rm int}$. We will
choose
\begin{align}
U_1 = K, ~~ U_2 = -1 + K^2,
\label{Eq:Cancellation_Condition}
\end{align}
so that the interactions in $S_{\rm bfg}[v,\chi]$ and in $S_{\rm
int}[v,\chi]$ exactly cancel off each other.

For every fixed configuration of the vector field $v_n$, $S_{\rm
bfg} + S_{\rm int}$ contains one and only one propagating Majorana
fermion field
\begin{align}
\xi_n \equiv \sqrt{ K } \sum_i \chi_{n,i} v_{n,i}.
\label{Eq:Fermion_Rewriting}
\end{align}
There are also two other Majorana fermion fields, denoted as $\xi'_n$ and $\xi''_n$, orthogonal to $\xi_n$ in the
$\SO(3)$ ``color space". For a fixed configuration of the vector boson $v_n$, we can perform a basis rotation from the $\chi_n$ field to $\xi_n$, $\xi_n'$ and $\xi_n''$ in the path integral. Notice that the fermion fields $\xi_n'$ and $\xi_n''$ only have local mass terms but not any kinetic terms (i.e. the $(\sigma^\mu-R)$ term). They can
be directly integrated out producing a factor that only depends on
the mass parameter $M$. Therefore, after integrating out $\xi_n'$ and $\xi_n''$ , we can write 
\begin{align}
& \int D[O^{n\mu}] \int D[\chi_{n,i}] e^{-S_{\rm bg} - S_{\rm fg} - S_{\rm int }}
\nonumber \\
& = \mathcal{N} \int D[\xi_n]
\exp\left(-\sum_{n,\mu} \bar{\xi}_{n+\mu}  (\sigma^\mu -R) \xi_n  - \sum_n M' \bar{\xi}_{n} \xi_{n}\right),
\label{Eq:Dual_Action_SO3}
\end{align}
where $\mathcal{N}$ is an overall normalization constant and the
mass parameter $M'$ for the Majorana fermion mode $\xi_n$ is given
by
\begin{align}
M' = M/K.
\end{align}
Now, we've arrived at the dual theory which exactly describes free
Majorana fermions on the Euclidean spacetime lattice. Although Eq.
\ref{Eq:Dual_Action_SO3} is derived on fixed configuration of the
boson field $v_n$, its right hand side does not depend on $v_n$. The independence on the vector boson configuration $v_n$ is expected also from the fact that any fixed vector boson configuration $v_n$ can be connected to any other by $\SO(3)$ gauge transformations before we perform any integration on the left hand side. Therefore, further integration over the boson field $v_n$ in Eq. \ref{Eq:Dual_Action_SO3} will
only introduce an extra overall multiplicative factor and will not
change the nature of the duality. At this point, we have constructed
an exact mapping between the theory in Eq.
\ref{Eq:Lattice_Boson_Fermion_theory} and free Majorana fermions
described by the action in Eq. \ref{Eq:Dual_Action_SO3}. This
exact mapping is valid regardless of whether or not the model
$Z_b$ (and its dual Majorana fermion theory) is at the critical
point or not.

Having already set $R=-1$, if we further take $M'=3$, the dual
Majorana fermion $\xi_n$, based on the change of Chern number for $M'>3$ and $M'<3$ givein in Eq.
\ref{Eq:Majorana_Chern_Number}, is exactly at a critical point
described by a single free gapless Majorana fermion in the IR. It implies
that the model Eq. \ref{Eq:Lattice_Boson_Fermion_theory} with the
following choices of parameter is also critical:
\begin{align}
M = 3 K, ~~~ U_1 = K, ~~  U_2 = -1 + K^2 .
\label{Eq:critical_coupling}
\end{align}
Now, we would like to return to the discussion of the continuum limit
of the model Eq. \ref{Eq:Lattice_Boson_Fermion_theory} at the
criticality. First of all, $0<K<1$ for any positive coupling $J$.
Hence, $M$ is always smaller than 3 at the critical point. If we choose $J$ such that
$1<M<3$ and neglect the effect of the interaction in $S_{\rm int}$
as we first integrate out the fermions $\chi_n$ in the model Eq.
\ref{Eq:Lattice_Boson_Fermion_theory}, we would naturally identify
the resulting model as a $\SO(3)$ critical bosons coupled to an
$\SO(3)_1$ Chern-Simons gauge theory. The presence of finite $U_{1,2}$
complicates the integration over the fermions field $\chi_n$ in
Eq. \ref{Eq:Lattice_Boson_Fermion_theory}. However, in a naive
continuum limit of Eq. \ref{Eq:Lattice_Boson_Fermion_theory}, the
interactions appear in $S_{\rm int}$ all contain high powers of
spacetime derivatives and may be considered irrelevant in a continuum
field theory. Furthermore, when the deviation of $M$ from 3 (as
well as 1) where the bare mass of the Majorana fermion $\chi_n$ vanishes,
is of order 1, the bare mass of the Majorana fermion $\chi_n$ in
Eq. \ref{Eq:Lattice_Boson_Fermion_theory} is of order 1. Although
Eq. \ref{Eq:critical_coupling} will require $U_{1,2}$ to be also
of order 1 at the same time, the small prefactor in $S_{\rm int}$
in defining the coupling constant, as well as the high power of
spacetime derivatives, may only lead to a small correction to the
integration over $\chi_n$ in Eq.
\ref{Eq:Lattice_Boson_Fermion_theory} compared to the case with
vanishing $U_{1,2}$. Since the level of the Chern-Simons term has
to be quantized, corrections induced by the interaction $S_{\rm
int}$, if small, will not be able to change it. Hence, we
speculate that the Euclidean spacetime lattice path integral
Eq.~\ref{Eq:Lattice_Boson_Fermion_theory} still corresponds to the
field theory Eq.~\ref{Eq:BosonContinuumTheory} in the IR with $g$
flows to a fixed point value.

\subsection{Correlation Functions and Global Symmetry}
\label{Sec:Global_Sym}
Regardless of the interpretation of the theory Eq.
\ref{Eq:Lattice_Boson_Fermion_theory} in the continuum limit, the
mapping discussed above is explicit and exact, which allow us to
identify not only the partition functions on both sides of the
duality but also the correlation functions. In particular, any
correlation functions of the Majorana fermion $\xi_n$ can be
exactly reproduced in the model Eq.
\ref{Eq:Lattice_Boson_Fermion_theory} by the operators defined in
Eq. \ref{Eq:Fermion_Rewriting}.

The exact mapping also helps us keep track of the global symmetry
on the two sides of the duality. The model Eq.
\ref{Eq:Lattice_Boson_Fermion_theory} possesses a global $Z_2$
symmetry:
\begin{align}
Z_2: v_n \rightarrow - v_n,~~~ \chi_n \rightarrow -\chi_n,
\end{align}
while it is evident from Eq. \ref{Eq:Fermion_Rewriting} that the free
Majorana fermion $\xi_n$ is neutral under this $Z_2$ symmetry.
This apparent mismatch of the $Z_2$ global symmetry on the two
lattice theories raises an potential subtlety in the IR duality of
their speculated continuum counterparts Eq.
\ref{Eq:BosonContinuumTheory} and Eq.
\ref{Eq:FermionContinuumTheory}, where the $Z_2$ symmetry in the
continuum field theories acts as $Z_2: \phi \rightarrow -\phi,~ \xi
\rightarrow  \xi$. This subtlety is not induced by the lattice
construction and was already noticed in Ref.
\onlinecite{Metlitski2017} for the continuum field theories. The
conjectured resolution of this symmetry mismatch is that the $Z_2$
symmetry is not spontaneously broken and the energy gap of the
$Z_2$ charged excitations remains finite across the critical point
on the boson side of the duality.

If we assume that the lattice theory Eq.
\ref{Eq:Lattice_Boson_Fermion_theory} correctly captures in the
continuum boson theory Eq. \ref{Eq:BosonContinuumTheory} in the
IR, we can direct test the aforementioned conjecture. In the
lattice theory Eq. \ref{Eq:Lattice_Boson_Fermion_theory}, there are two $\SO(3)$
gauge-invariant $Z_2$-charged (local) operators: 
\begin{align}
\epsilon_{ii'i''} v_{n,i} \chi_{n,i'}
\chi_{n,i''}~~\text{and}~~\epsilon_{ii'i''} \chi_{n,i} \chi_{n,i'}
\chi_{n,i''},
\end{align}
where we have suppressed the spinor indices of the fermion field
$\chi_n$. Being an bosonic operator that is $\SO(3)$
gauge-invariant and odd under the $Z_2$ symmetry,
$\epsilon_{ii'i''} v_{n,i} \chi_{n,i'} \chi_{n,i''}$ can serve as
an order parameter for any potential spontaneous breaking of the
global $Z_2$ symmetry in the theory Eq.
\ref{Eq:Lattice_Boson_Fermion_theory}. Interestingly, we notice
that the operator $\epsilon_{ii'i''} v_{n,i} \chi_{n,i'}
\chi_{n,i''}$ is always proportional to the product of the fermion
fields $\xi_n'$ and $\xi_n''$ that are introduced after we
integrate out the $\SO(3)$ gauge field in Eq.
\ref{Eq:Eff_Action_From_SO3} and that are orthogonal to the
low-energy fermion field $\xi_n$ defined in Eq.
\ref{Eq:Fermion_Rewriting}. As stated previously, unlike the
fermion field $\xi_n$, the fields $\xi_n'$ and $\xi_n''$ only have
local mass terms but no kinetic term. Therefore, any correlation
of the fields $\xi_n'$ and $\xi_n''$ is short-range, which implies
that the correlation of the operator $\epsilon_{ii'i''} v_{n,i}
\chi_{n,i'} \chi_{n,i''}$ is also short-range and, consequently,
that the $Z_2$ global symmetry is unbroken. This statement is in
agreement with Ref. \onlinecite{Metlitski2017}. Furthermore, it is
evident directly from the derivation of the exact mapping in the
previous subsection, that the only low-energy field at the
critical point is the fermion field $\xi_n$ which is neutral under
the global symmetry $Z_2$. That in turn implies that all the $Z_2$
charged excitations in the lattice theory Eq.
\ref{Eq:Lattice_Boson_Fermion_theory} are gapped across the
critical point. This observation offers another strong evidence
for the conjecture on the $Z_2$ global symmetry proposed in Ref.
\onlinecite{Metlitski2017}.

\subsection{An Alternative Dual Model}
\label{Sec:Alternative_Dual}
In Sec. \ref{Sec:ExactMap}, we enforce the condition Eq. \ref{Eq:Cancellation_Condition} to ensure that the dual Majorana fermion $\xi_n$ is exactly free. In this subsection, we proceed with the lattice duality without this condition Eq. \ref{Eq:Cancellation_Condition}. In fact, for generic values of $U_{1,2}$ with $U_1 < K$, the path integral $e^{-S_{\rm bfg}-S_{\rm int}}$ (in a fixed background of vector boson configuration $v_n$) can be viewed as containing an $\SO(2)$ gauge field in disguise on the dual side (see Appendix \ref{App:Derivation_Dual}). We will discuss the physical meaning of this $\SO(2)$ gauge field later. To elucidate the $\SO(2)$ gauge structure, we first introduce an $\SO(2)$ gauge field represented by a $2\times 2$ orthogonal matrix $U^{n\mu}_{ab}$ (with $a,b=1,2$) on each link (connecting site $n$ and site $n+\mu$). Then, we construct the Majorana field $\eta_{n,a}$ that is charged under the $\SO(2)$ gauge field:
\begin{align}
\eta_{n,1} =  (K- U_1)^{\frac{1}{4}} \sum_i w_{n,i}' \chi_{n,},
\nonumber \\
\eta_{n,2} =  (K- U_1)^{\frac{1}{4}}  \sum_i w_{n,i}'' \chi_{n,i},
\label{Eq:eta_fermion_Def}
\end{align}
where $w_{n}'$ and $w_{n}''$ are two mutually orthogonal 3-component unit vectors that are both orthogonal to the fixed vector bosons field background $v_n$. The subscript $a$ in $\eta_{n,a}$ represents the SO(2) ``color" index. In fact, the fermion fields $\eta_{n,a}$ are essential the same as the fermion fields $\xi_n'$ and $\xi_n''$ discussed in Sec. \ref{Sec:ExactMap} but with an extra $(K-U_1)^{\frac{1}{4}}$ prefactor that makes the fields $\eta_{n,a}$ well-defined only when $K>U_1$. As we will see, unlike the non-propagating fermion fields $\xi_n'$ and $\xi_n''$ in Sec. \ref{Sec:ExactMap}, $\eta_{n,a}$ can be interpreted as propagating fermion fields in dual theory thanks to $K>U_1$. Having introduced these ingredients, we can rewrite $e^{-S_{\rm bfg}-S_{\rm int}}$ as (see Appendix \ref{App:Derivation_Dual} for detailed derivation)
\begin{widetext}
\begin{align}
 e^{- S_{\rm bfg} - S_{\rm int}} &  = 
\exp \left[- \left( \sum_{n}  M' \bar{\xi}_{n} \xi_{n} + \sum_{n,\mu} \bar{\xi}_{n+\mu}  (\sigma^\mu -R) \xi_{n}  \right)   \right] 
\nonumber 
\\
& \times 
\int D[U^{n\mu}] 
\exp\left[-\left(\sum_n M'' \bar{\eta}_{n,a} \eta_{n,a}  +\sum_{n,\mu}\bar{\eta}_{n+\mu,a}  (\sigma^\mu -R) U^{n\mu}_{ab} \eta_{n,b} \right) \right] 
\nonumber 
\\
&
\times \exp \left[ 
-V \sum_{n,\mu}
\left( \bar{\xi}_{n+\mu}  (\sigma^\mu -R) \xi_{n} \right)
  \left( \bar{\eta}_{n+\mu,1}  (\sigma^\mu -R) \eta_{n,1} \right) 
  \left( \bar{\eta}_{n+\mu,2}  (\sigma^\mu -R) \eta_{n,2} \right) 
\right],
\label{Eq:Alternative_Dual}
\end{align}
\end{widetext}
where the repeated $\SO(2)$ indices $a,b$ are automatically summed from 1 to 2. The integration of the SO(2) gauge configuration $\int D[U^{n\mu}] $ should be understood as the integration of the matrix $U^{n\mu}$ over the Haar measure of $\SO(2)$ on every link of the spacetime lattice. The mass parameters $M'$, $M''$ of the fermion $\xi_n$ and $\eta_n$ are given by
\begin{align}
M' = M/K, ~~ M'' = M (K- U_1)^{-\frac{1}{2}}, 
\end{align}  
and the coupling constant $V$ by
\begin{align}
V= \frac{1+U_2  -K^2}{(K-U_1)K}
\end{align}
We can see that the right hand side of Eq. \ref{Eq:Alternative_Dual} is a theory that describes a Majorana fermion $\xi_n$ and Majorana fermions $\eta_{n,a}$ coupled to an $\SO(2)$ gauge field. The two types of fermions interact with each other via the $\SO(2)$ gauge-invariant interaction given in the third line of Eq. \ref{Eq:Alternative_Dual}.

Physically, the $\SO(2)$ gauge field introduced in the dual theory in Eq. \ref{Eq:Alternative_Dual} essentially recovers the residue $\SO(2)$ subgroup of the $\SO(3)$ gauge group in the original model Eq. \ref{Eq:Lattice_Boson_Fermion_theory} in the presence of a fixed vector boson configuration $v_n$. The way to understand it is to notice that the fermions $\eta_{n,1}$ and $\eta_{n,2}$ that are charged under the $\SO(2)$ gauge field are essentially the two fermion fields orthogonal to $\xi_n$ (as well as the fixed vector field value $v_n$) is the $\SO(3)$ color space. The $\SO(2)$ gauge group can be viewed as the residue gauge group of $\SO(3)$ that preserves the vector field configuration $v_n$. As we tune to the parameter regime where $M'$ is close to 3, $K$ is close to $U_1$ and $V$ is small. The Majorana fermion $\chi_n$ will have a small bare mass, while the mass parameter $M''$ of the Majorana fermion $\eta_n$ can be set to be $M''\gg 3$ which ensures not only a big energy gap but also a trivial bare band stucture of $\eta_n$. Furthermore, if $V$ is tuned to a small value by tuning $U_2$, the dual theory will only have an ``almost free" (or weakly-interacting) Majorana fermion $\xi_n$ at low energy. Since the dual low-energy Majorana fermion $\xi_n$ is now weakly interacting, the exact position of the critical point of the dual theory Eq. \ref{Eq:Alternative_Dual} (as well as the original model Eq. \ref{Eq:Lattice_Boson_Fermion_theory}) is expected to be shifted from $M'=3$.

\section{Lattice Duality with O(3) gauge fields}
\label{Sec:O(3)Duality} In this section, we generalize the model
Eq. \ref{Eq:Lattice_Boson_Fermion_theory} studied in the
previous section to the case with $\O(3)$ gauge fields. The action
$S_{\rm bg}[v, O]$ given in Eq. \ref{Eq: Action_SO(3)_vectors} and
$S_{\rm fg}[\chi, O]$ in Eq. \ref{Eq: Action_SO(3)_Majorana} can
be directly promoted to accommodate $\O(3)$ gauge field
$O^{n\mu}$. We will rename them as $S_{\rm bg}'[v, O]$ and $S_{\rm
fg}'[\chi, O]$ in order to distinguish them from their
$\SO(3)$-gauge-group counterpart. All the interactions in $S_{\rm
int}$ are not invariant under the $\O(3)$ gauge transforms. Although one can always consider a ``O(3)-gauged" version of $S_{\rm int}$, we will exclude them from the generalized model for simplicity. The
Euclidean lattice path integral of the generalized model is given
by
\begin{align}
Z_b' = \int D[O^{n\mu}] \int D[\chi_{n,i}] \int D[v_n] e^{-S'_{\rm
bg} - S'_{\rm fg}}, \label{Eq:Lattice_Boson_Fermion_theory_O3}
\end{align}
where the integration of the boson field $v_n$ and the fermion
field $\chi_n$ follow the same rules as before. The integration of
the gauge field $\int D[O^{n\mu}]$ is now carried over the Haar
measure of $\O(3)$ instead of $\SO(3)$. However, since we can
write $\O(3) = \SO(3)\times Z_2$, the integration over $\O(3)$ can
be viewed as the integration over the Haar measure of $\SO(3)$
together with an ``integration" over the $Z_2$ subgroup.

Similar to the treatment of the previous section, we can first
integrate out the $\O(3)$ gauge field
\begin{align}
& \int D[O^{n\mu}] ~ e^{-S'_{\rm bg}} e^{ - S'_{\rm fg}} =
\left(\frac{2\sinh J}{J}\right)^{3N_s} e^{-S'_{\rm bfg}[v,\chi]}
\end{align}
where the effective action $S'_{\rm bfg}[v,\chi]$ takes the form
\begin{align}
& S'_{\rm bfg}[v,\chi] =
\nonumber
\\
&  \sum_{n}  M \bar{\chi}_{n,i} \chi_{n,i}  + \sum_{n,\mu}  K
(\bar{\chi}_{n+\mu,i} v_{n+\mu,i}) (\sigma^\mu -R) (\chi_{n,j}
v_{n,j}), \label{Eq:Eff_Action_From_O3}
\end{align}
which is much simpler compared to Eq. \ref{Eq:Eff_Action_From_SO3}
of the $\SO(3)$ case. Again, we introduce the fermion field
$\xi_n$ given in Eq. \ref{Eq:Fermion_Rewriting} and integrate out
the orthogonal fermion fields $\xi_n'$ and $\xi_n''$ to obtain the
following exact mapping
\begin{align}
& \int D[O^{n\mu}] ~ e^{-S'_{\rm bg}} e^{ - S'_{\rm fg}}  =
\nonumber \\
& \mathcal{N}' \int D[\xi] \exp\left(-\sum_{n,\mu}
\bar{\xi}_{n+\mu}  (\sigma^\mu -R) \xi_n  - \sum_n M'
\bar{\xi}_{n} \xi_{n}\right), \label{Eq:Dual_Fermion_O3}
\end{align}
where $\mathcal{N}'$ is an overall normalization constant and the
mass parameter $M'$ given by
\begin{align}
M' = M/K.
\end{align}

Now, we've established an exact mapping from the model Eq.
\ref{Eq:Lattice_Boson_Fermion_theory_O3} to a model with the free
Majorana freemion $\xi_n$ on the dual side. Again, we emphasize that this
lattice duality is exact and is valid for any choice of coupling
constant $J$ and mass parameter $M$. Based on the exact mapping,
by setting $M'=M/K = 3$, we can tune both sides of the duality to
the critical point where one side of the duality is captured by a
massless free Majorana fermion in the IR.

While one side of the duality is naturally identified with the free
massless Majorana fermion in the continuum limit, the IR nature
of the model Eq. \ref{Eq:Lattice_Boson_Fermion_theory_O3} remains
unclear. One may follow the same strategy as the previous section
to integrate over the fermion field $\chi$ first and interpret the
continuum theory as a vector boson coupled to a Chern-Simons term.
However, it is unclear how the $Z_2$ part of the $\O(3)$ gauge
field behaves after we integrate out the fermions field $\chi_n$.
A possible reconciliation with the duality studied in the previous
section and the IR duality between the continuum field theories
Eq. \ref{Eq:BosonContinuumTheory} and Eq.
\ref{Eq:FermionContinuumTheory} is that only the $\SO(3)$ subgroup
of the $\O(3)$ is deconfined. Another puzzling feature of this
lattice duality with $\O(3)$ gauge field is that while the dual
Majorana fermion stays at the critical point, we can freely tune
the parameters $K$ such that the bare band structure of the
fermions $\chi_n$ in the model Eq.
\ref{Eq:Lattice_Boson_Fermion_theory_O3} can have different Chern
numbers (and the naively expected Chern-Simons levels): $C=1$ when $1<M<3$ and $C=-2$ when $M<1$. We will defer
resolution of these puzzles to future studies.

Similar boson-fermion lattice duality with different gauge groups can also be constructed following the same method we discussed here. For example, as detailed in Appendix \ref{sec:Z2 gauge}, when the gauge group is reduced to $Z_2$, we can obtain a lattice duality between a free massless Majorana fermion and a $Z_2$ matter field coupled to a $Z_2$ gauge field.

The lattice duality constructed in this section has an $\O(3)$ gauge field on one side and a single Majorana fermion on the other. Interestingly, Ref. \onlinecite{Cordova2017} propose an IR duality with similar but yet distinct features. The proposed IR duality is between a critical vector boson coupled to an $\O(3)_{1,1}^0$ Chern-Simons gauge theory (following Ref. \onlinecite{Cordova2017}'s notation) and a single Majorana fermion plus a decoupled $Z_2$ gauge theory. In fact, as detailed in Appendix \ref{App:Majorana_Z2}, a modified version of the model Eq. \ref{Eq:Lattice_Boson_Fermion_theory_O3}, can be exactly mapped to a free Majorana fermion plus a decoupled $Z_2$ gauge theory on the lattice. The modified model is on its own is speculated to be connected to in the IR the critical vector boson coupled to an $\O(3)_{1,1}^0$ Chern-Simons gauge theory.

\section{Summary and Discussion}

In this work, we construct the Euclidean spacetime lattice path
integral of a theory with strongly interacting bosons and Majorana
fermions coupled to $\SO(3)$ gauge field and exactly map this
theory to the theory of free Majorana fermions. This exact mapping
or lattice duality is argued to be connected to the IR duality
between critical bosons coupled to a $\SO(3)_1$ Chern-Simons term
and a single free Majorana fermion proposed in
Ref.~\onlinecite{Metlitski2017,Hsin2017}. This lattice duality
provides an exact mapping of the Euclidean path integrals and
correlation functions between both sides of the duality, which
allows us to obtain a strong evidence for the conjecture in Ref.
\onlinecite{Metlitski2017} regarding the apparent mismatch of a
global $Z_2$ symmetry on the two sides of the IR duality. This
model is also generalized to lattice theories with $\O(3)$ gauge
fields and with $Z_2$ gauge fields respectively, which are both again exactly dualized to a free Majorana fermion. A different generalization of the lattice model with $\O(3)$ gauge fields is shown to be dual to a free Majorana fermion plus a decoupled $Z_2$ gauge field.

While the lattice dualities are exact, the fate of these strongly
interacting lattice theories in the IR still needs more attention.
In Sec. \ref{Sec:SO(3)Duality}, we argue that the model Eq.
\ref{Eq:Lattice_Boson_Fermion_theory} with strongly-interacting
boson and fermions coupled to $\SO(3)$ gauge field should be
related to the critical boson coupled to a Chern-Simons term in
the IR. To solidify this argument, we need to understand the
effect of $S_{\rm int}$, in particular, on the Chern-Simons level
when we integrate out the fermion fields $\chi_n$ in Eq.
\ref{Eq:Lattice_Boson_Fermion_theory}. Knowing that the gauge
invariant fermion in the lattice duality has a critical point, the
bosonic side of the duality should certainly also have a critical
point, but whether this critical point really corresponds to the
one described by the field theory
Eq.~\ref{Eq:BosonContinuumTheory} is difficult to prove. This is
certainly a question to be answered in the future. Similarly, as
we discussed in Sec. \ref{Sec:O(3)Duality}, the IR nature of the
theory of strongly-interacting boson and fermions coupled to the
$\O(3)$ gauge field given in Eq.
\ref{Eq:Lattice_Boson_Fermion_theory_O3} also remains unclear. In
particular, the role of the $Z_2$ subgroup of the $\O(3)$ gauge
group and the Chern number of the bare band structure of the
fermion fields $\chi_n$ both require further investigation.

In a broader picture, it is an interesting direction to generalize
the lattice construction of dualities to interacting theories with
more general non-Abelian gauge groups and more flavors of
interacting bosons or fermions. While being exact without the use
of any large-$N$ limit, holography or supersymmetry, the lattice
construction is potentially a powerful tool in deriving previously
proposal IR dualities with non-Abelian gauge fields as well as
discovering new ones. Once the lattice construction of these
duality is constructed, one should be able to read out the
correspondence between the operators on both sides of the duality.

{\it Note added:} After the completion of this work, we learned of
Ref. \onlinecite{ChenDual2018}, which is a similar and independent attempt to
generalize the lattice construction to tackle dualities with
non-Abelian gauge fields.

\begin{acknowledgements}

The authors thank Cenke Xu for insightful discussions. CMJ thanks Jing-Yuan Chen for private communication on Ref. \onlinecite{ChenDual2018} prior to its arXiv posting. ZB acknowledges support from the Pappalardo fellowship at MIT.

\end{acknowledgements}

\bibliography{TI}

\begin{thebibliography}{56}
\expandafter\ifx\csname natexlab\endcsname\relax\def\natexlab#1{#1}\fi
\expandafter\ifx\csname bibnamefont\endcsname\relax
  \def\bibnamefont#1{#1}\fi
\expandafter\ifx\csname bibfnamefont\endcsname\relax
  \def\bibfnamefont#1{#1}\fi
\expandafter\ifx\csname citenamefont\endcsname\relax
  \def\citenamefont#1{#1}\fi
\expandafter\ifx\csname url\endcsname\relax
  \def\url#1{\texttt{#1}}\fi
\expandafter\ifx\csname urlprefix\endcsname\relax\def\urlprefix{URL }\fi
\providecommand{\bibinfo}[2]{#2}
\providecommand{\eprint}[2][]{\url{#2}}

\bibitem[{\citenamefont{Peskin}(1978)}]{Peskin1978}
\bibinfo{author}{\bibfnamefont{M.~E.} \bibnamefont{Peskin}},
  \bibinfo{journal}{Annals of Physics} \textbf{\bibinfo{volume}{113}},
  \bibinfo{pages}{122 } (\bibinfo{year}{1978}), ISSN \bibinfo{issn}{0003-4916}.

\bibitem[{\citenamefont{Dasgupta and Halperin}(1981)}]{Dasgupta1981}
\bibinfo{author}{\bibfnamefont{C.}~\bibnamefont{Dasgupta}} \bibnamefont{and}
  \bibinfo{author}{\bibfnamefont{B.~I.} \bibnamefont{Halperin}},
  \bibinfo{journal}{Phys. Rev. Lett.} \textbf{\bibinfo{volume}{47}},
  \bibinfo{pages}{1556} (\bibinfo{year}{1981}).

\bibitem[{\citenamefont{{Polyakov}}(1988)}]{Polyakov1988}
\bibinfo{author}{\bibfnamefont{A.~M.} \bibnamefont{{Polyakov}}},
  \bibinfo{journal}{Modern Physics Letters A} \textbf{\bibinfo{volume}{3}},
  \bibinfo{pages}{325} (\bibinfo{year}{1988}).

\bibitem[{\citenamefont{Chen et~al.}(1993)\citenamefont{Chen, Fisher, and
  Wu}}]{wufisher}
\bibinfo{author}{\bibfnamefont{W.}~\bibnamefont{Chen}},
  \bibinfo{author}{\bibfnamefont{M.~P.~A.} \bibnamefont{Fisher}},
  \bibnamefont{and} \bibinfo{author}{\bibfnamefont{Y.-S.} \bibnamefont{Wu}},
  \bibinfo{journal}{Phys. Rev. B} \textbf{\bibinfo{volume}{48}},
  \bibinfo{pages}{13749} (\bibinfo{year}{1993}).

\bibitem[{\citenamefont{Giombi et~al.}(2012)\citenamefont{Giombi, Minwalla,
  Prakash, Trivedi, Wadia, and Yin}}]{Giombi2012}
\bibinfo{author}{\bibfnamefont{S.}~\bibnamefont{Giombi}},
  \bibinfo{author}{\bibfnamefont{S.}~\bibnamefont{Minwalla}},
  \bibinfo{author}{\bibfnamefont{S.}~\bibnamefont{Prakash}},
  \bibinfo{author}{\bibfnamefont{S.~P.} \bibnamefont{Trivedi}},
  \bibinfo{author}{\bibfnamefont{S.~R.} \bibnamefont{Wadia}}, \bibnamefont{and}
  \bibinfo{author}{\bibfnamefont{X.}~\bibnamefont{Yin}}, \bibinfo{journal}{The
  European Physical Journal C} \textbf{\bibinfo{volume}{72}},
  \bibinfo{pages}{2112} (\bibinfo{year}{2012}), ISSN \bibinfo{issn}{1434-6052},
  \urlprefix\url{https://doi.org/10.1140/epjc/s10052-012-2112-0}.

\bibitem[{\citenamefont{{Aharony}
  et~al.}(2012{\natexlab{a}})\citenamefont{{Aharony}, {Gur-Ari}, and
  {Yacoby}}}]{Aharony2012_1}
\bibinfo{author}{\bibfnamefont{O.}~\bibnamefont{{Aharony}}},
  \bibinfo{author}{\bibfnamefont{G.}~\bibnamefont{{Gur-Ari}}},
  \bibnamefont{and} \bibinfo{author}{\bibfnamefont{R.}~\bibnamefont{{Yacoby}}},
  \bibinfo{journal}{Journal of High Energy Physics}
  \textbf{\bibinfo{volume}{12}}, \bibinfo{eid}{28}
  (\bibinfo{year}{2012}{\natexlab{a}}), \eprint{1207.4593}.

\bibitem[{\citenamefont{{Aharony}
  et~al.}(2012{\natexlab{b}})\citenamefont{{Aharony}, {Gur-Ari}, and
  {Yacoby}}}]{Aharony2012_2}
\bibinfo{author}{\bibfnamefont{O.}~\bibnamefont{{Aharony}}},
  \bibinfo{author}{\bibfnamefont{G.}~\bibnamefont{{Gur-Ari}}},
  \bibnamefont{and} \bibinfo{author}{\bibfnamefont{R.}~\bibnamefont{{Yacoby}}},
  \bibinfo{journal}{Journal of High Energy Physics}
  \textbf{\bibinfo{volume}{3}}, \bibinfo{eid}{37}
  (\bibinfo{year}{2012}{\natexlab{b}}), \eprint{1110.4382}.

\bibitem[{\citenamefont{Seiberg et~al.}(2016)\citenamefont{Seiberg, Senthil,
  Wang, and Witten}}]{seiberg1}
\bibinfo{author}{\bibfnamefont{N.}~\bibnamefont{Seiberg}},
  \bibinfo{author}{\bibfnamefont{T.}~\bibnamefont{Senthil}},
  \bibinfo{author}{\bibfnamefont{C.}~\bibnamefont{Wang}}, \bibnamefont{and}
  \bibinfo{author}{\bibfnamefont{E.}~\bibnamefont{Witten}},
  \bibinfo{journal}{Annals of Physics} \textbf{\bibinfo{volume}{374}},
  \bibinfo{pages}{395 } (\bibinfo{year}{2016}), ISSN \bibinfo{issn}{0003-4916}.

\bibitem[{\citenamefont{{Thanh Son}}(2015)}]{Son2015}
\bibinfo{author}{\bibfnamefont{D.}~\bibnamefont{{Thanh Son}}},
  \bibinfo{journal}{ArXiv e-prints}  (\bibinfo{year}{2015}),
  \eprint{1502.03446}.

\bibitem[{\citenamefont{{Metlitski} and
  {Vishwanath}}(2016)}]{MetlitskiDualDirac}
\bibinfo{author}{\bibfnamefont{M.~A.} \bibnamefont{{Metlitski}}}
  \bibnamefont{and}
  \bibinfo{author}{\bibfnamefont{A.}~\bibnamefont{{Vishwanath}}},
  \bibinfo{journal}{\prb} \textbf{\bibinfo{volume}{93}}, \bibinfo{eid}{245151}
  (\bibinfo{year}{2016}), \eprint{1505.05142}.

\bibitem[{\citenamefont{{Wang} and {Senthil}}(2015)}]{WangDualDirac}
\bibinfo{author}{\bibfnamefont{C.}~\bibnamefont{{Wang}}} \bibnamefont{and}
  \bibinfo{author}{\bibfnamefont{T.}~\bibnamefont{{Senthil}}},
  \bibinfo{journal}{Physical Review X} \textbf{\bibinfo{volume}{5}},
  \bibinfo{eid}{041031} (\bibinfo{year}{2015}), \eprint{1505.05141}.

\bibitem[{\citenamefont{Mross et~al.}(2016)\citenamefont{Mross, Alicea, and
  Motrunich}}]{mrossduality}
\bibinfo{author}{\bibfnamefont{D.~F.} \bibnamefont{Mross}},
  \bibinfo{author}{\bibfnamefont{J.}~\bibnamefont{Alicea}}, \bibnamefont{and}
  \bibinfo{author}{\bibfnamefont{O.~I.} \bibnamefont{Motrunich}},
  \bibinfo{journal}{Phys. Rev. Lett.} \textbf{\bibinfo{volume}{117}},
  \bibinfo{pages}{016802} (\bibinfo{year}{2016}).

\bibitem[{\citenamefont{{Wang} et~al.}(2017)\citenamefont{{Wang}, {Nahum},
  {Metlitski}, {Xu}, and {Senthil}}}]{WangO5}
\bibinfo{author}{\bibfnamefont{C.}~\bibnamefont{{Wang}}},
  \bibinfo{author}{\bibfnamefont{A.}~\bibnamefont{{Nahum}}},
  \bibinfo{author}{\bibfnamefont{M.~A.} \bibnamefont{{Metlitski}}},
  \bibinfo{author}{\bibfnamefont{C.}~\bibnamefont{{Xu}}}, \bibnamefont{and}
  \bibinfo{author}{\bibfnamefont{T.}~\bibnamefont{{Senthil}}},
  \bibinfo{journal}{Physical Review X} \textbf{\bibinfo{volume}{7}},
  \bibinfo{eid}{031051} (\bibinfo{year}{2017}), \eprint{1703.02426}.

\bibitem[{\citenamefont{{Seiberg} et~al.}(2016)\citenamefont{{Seiberg},
  {Senthil}, {Wang}, and {Witten}}}]{SSWW2016}
\bibinfo{author}{\bibfnamefont{N.}~\bibnamefont{{Seiberg}}},
  \bibinfo{author}{\bibfnamefont{T.}~\bibnamefont{{Senthil}}},
  \bibinfo{author}{\bibfnamefont{C.}~\bibnamefont{{Wang}}}, \bibnamefont{and}
  \bibinfo{author}{\bibfnamefont{E.}~\bibnamefont{{Witten}}},
  \bibinfo{journal}{Annals of Physics} \textbf{\bibinfo{volume}{374}},
  \bibinfo{pages}{395} (\bibinfo{year}{2016}), \eprint{1606.01989}.

\bibitem[{\citenamefont{{Hsin} and {Seiberg}}(2016)}]{Hsin2016}
\bibinfo{author}{\bibfnamefont{P.-S.} \bibnamefont{{Hsin}}} \bibnamefont{and}
  \bibinfo{author}{\bibfnamefont{N.}~\bibnamefont{{Seiberg}}},
  \bibinfo{journal}{Journal of High Energy Physics}
  \textbf{\bibinfo{volume}{9}}, \bibinfo{eid}{95} (\bibinfo{year}{2016}),
  \eprint{1607.07457}.

\bibitem[{\citenamefont{{Xu} and {You}}(2015)}]{YouXu2015}
\bibinfo{author}{\bibfnamefont{C.}~\bibnamefont{{Xu}}} \bibnamefont{and}
  \bibinfo{author}{\bibfnamefont{Y.-Z.} \bibnamefont{{You}}},
  \bibinfo{journal}{\prb} \textbf{\bibinfo{volume}{92}}, \bibinfo{eid}{220416}
  (\bibinfo{year}{2015}), \eprint{1510.06032}.

\bibitem[{\citenamefont{{Aharony} et~al.}(2017)\citenamefont{{Aharony},
  {Benini}, {Hsin}, and {Seiberg}}}]{Hsin2017}
\bibinfo{author}{\bibfnamefont{O.}~\bibnamefont{{Aharony}}},
  \bibinfo{author}{\bibfnamefont{F.}~\bibnamefont{{Benini}}},
  \bibinfo{author}{\bibfnamefont{P.-S.} \bibnamefont{{Hsin}}},
  \bibnamefont{and}
  \bibinfo{author}{\bibfnamefont{N.}~\bibnamefont{{Seiberg}}},
  \bibinfo{journal}{Journal of High Energy Physics}
  \textbf{\bibinfo{volume}{2}}, \bibinfo{eid}{72} (\bibinfo{year}{2017}),
  \eprint{1611.07874}.

\bibitem[{\citenamefont{{Aharony}}(2016)}]{Aharony2016}
\bibinfo{author}{\bibfnamefont{O.}~\bibnamefont{{Aharony}}},
  \bibinfo{journal}{Journal of High Energy Physics}
  \textbf{\bibinfo{volume}{2}}, \bibinfo{eid}{93} (\bibinfo{year}{2016}),
  \eprint{1512.00161}.

\bibitem[{\citenamefont{{Karch} and {Tong}}(2016)}]{Karch2016}
\bibinfo{author}{\bibfnamefont{A.}~\bibnamefont{{Karch}}} \bibnamefont{and}
  \bibinfo{author}{\bibfnamefont{D.}~\bibnamefont{{Tong}}},
  \bibinfo{journal}{Physical Review X} \textbf{\bibinfo{volume}{6}},
  \bibinfo{eid}{031043} (\bibinfo{year}{2016}), \eprint{1606.01893}.

\bibitem[{\citenamefont{{Metlitski} et~al.}(2017)\citenamefont{{Metlitski},
  {Vishwanath}, and {Xu}}}]{Metlitski2017}
\bibinfo{author}{\bibfnamefont{M.~A.} \bibnamefont{{Metlitski}}},
  \bibinfo{author}{\bibfnamefont{A.}~\bibnamefont{{Vishwanath}}},
  \bibnamefont{and} \bibinfo{author}{\bibfnamefont{C.}~\bibnamefont{{Xu}}},
  \bibinfo{journal}{\prb} \textbf{\bibinfo{volume}{95}}, \bibinfo{eid}{205137}
  (\bibinfo{year}{2017}), \eprint{1611.05049}.

\bibitem[{\citenamefont{{Gaiotto} et~al.}(2018)\citenamefont{{Gaiotto},
  {Komargodski}, and {Seiberg}}}]{Gaiotto2018}
\bibinfo{author}{\bibfnamefont{D.}~\bibnamefont{{Gaiotto}}},
  \bibinfo{author}{\bibfnamefont{Z.}~\bibnamefont{{Komargodski}}},
  \bibnamefont{and}
  \bibinfo{author}{\bibfnamefont{N.}~\bibnamefont{{Seiberg}}},
  \bibinfo{journal}{Journal of High Energy Physics}
  \textbf{\bibinfo{volume}{1}}, \bibinfo{eid}{110} (\bibinfo{year}{2018}),
  \eprint{1708.06806}.

\bibitem[{\citenamefont{{Gomis} et~al.}(2017)\citenamefont{{Gomis},
  {Komargodski}, and {Seiberg}}}]{Gomis2017}
\bibinfo{author}{\bibfnamefont{J.}~\bibnamefont{{Gomis}}},
  \bibinfo{author}{\bibfnamefont{Z.}~\bibnamefont{{Komargodski}}},
  \bibnamefont{and}
  \bibinfo{author}{\bibfnamefont{N.}~\bibnamefont{{Seiberg}}},
  \bibinfo{journal}{ArXiv e-prints}  (\bibinfo{year}{2017}),
  \eprint{1710.03258}.

\bibitem[{\citenamefont{{Karch} et~al.}(2017)\citenamefont{{Karch}, {Robinson},
  and {Tong}}}]{KarchRobinson2017}
\bibinfo{author}{\bibfnamefont{A.}~\bibnamefont{{Karch}}},
  \bibinfo{author}{\bibfnamefont{B.}~\bibnamefont{{Robinson}}},
  \bibnamefont{and} \bibinfo{author}{\bibfnamefont{D.}~\bibnamefont{{Tong}}},
  \bibinfo{journal}{Journal of High Energy Physics}
  \textbf{\bibinfo{volume}{1}}, \bibinfo{eid}{17} (\bibinfo{year}{2017}),
  \eprint{1609.04012}.

\bibitem[{\citenamefont{{Gur-Ari} and {Yacoby}}(2015)}]{GurAri2015}
\bibinfo{author}{\bibfnamefont{G.}~\bibnamefont{{Gur-Ari}}} \bibnamefont{and}
  \bibinfo{author}{\bibfnamefont{R.}~\bibnamefont{{Yacoby}}},
  \bibinfo{journal}{Journal of High Energy Physics}
  \textbf{\bibinfo{volume}{11}}, \bibinfo{eid}{13} (\bibinfo{year}{2015}),
  \eprint{1507.04378}.

\bibitem[{\citenamefont{{Benini}}(2018)}]{Benini2018}
\bibinfo{author}{\bibfnamefont{F.}~\bibnamefont{{Benini}}},
  \bibinfo{journal}{Journal of High Energy Physics}
  \textbf{\bibinfo{volume}{2}}, \bibinfo{eid}{68} (\bibinfo{year}{2018}),
  \eprint{1712.00020}.

\bibitem[{\citenamefont{{Jensen} and {Karch}}(2017)}]{Jensen2017}
\bibinfo{author}{\bibfnamefont{K.}~\bibnamefont{{Jensen}}} \bibnamefont{and}
  \bibinfo{author}{\bibfnamefont{A.}~\bibnamefont{{Karch}}},
  \bibinfo{journal}{Journal of High Energy Physics}
  \textbf{\bibinfo{volume}{11}}, \bibinfo{eid}{18} (\bibinfo{year}{2017}),
  \eprint{1709.01083}.

\bibitem[{\citenamefont{Jensen}(2018)}]{Jensen2018}
\bibinfo{author}{\bibfnamefont{K.}~\bibnamefont{Jensen}},
  \bibinfo{journal}{Journal of High Energy Physics}
  \textbf{\bibinfo{volume}{2018}}, \bibinfo{pages}{31} (\bibinfo{year}{2018}),
  ISSN \bibinfo{issn}{1029-8479},
  \urlprefix\url{https://doi.org/10.1007/JHEP01(2018)031}.

\bibitem[{\citenamefont{{Cordova}
  et~al.}(2017{\natexlab{a}})\citenamefont{{Cordova}, {Hsin}, and
  {Seiberg}}}]{Cordova2017_1}
\bibinfo{author}{\bibfnamefont{C.}~\bibnamefont{{Cordova}}},
  \bibinfo{author}{\bibfnamefont{P.-S.} \bibnamefont{{Hsin}}},
  \bibnamefont{and}
  \bibinfo{author}{\bibfnamefont{N.}~\bibnamefont{{Seiberg}}},
  \bibinfo{journal}{ArXiv e-prints}  (\bibinfo{year}{2017}{\natexlab{a}}),
  \eprint{1712.08639}.

\bibitem[{\citenamefont{{Cordova}
  et~al.}(2017{\natexlab{b}})\citenamefont{{Cordova}, {Hsin}, and
  {Seiberg}}}]{Cordova2017_2}
\bibinfo{author}{\bibfnamefont{C.}~\bibnamefont{{Cordova}}},
  \bibinfo{author}{\bibfnamefont{P.-S.} \bibnamefont{{Hsin}}},
  \bibnamefont{and}
  \bibinfo{author}{\bibfnamefont{N.}~\bibnamefont{{Seiberg}}},
  \bibinfo{journal}{ArXiv e-prints}  (\bibinfo{year}{2017}{\natexlab{b}}),
  \eprint{1711.10008}.

\bibitem[{\citenamefont{Cheng and Xu}(2016)}]{mengxu}
\bibinfo{author}{\bibfnamefont{M.}~\bibnamefont{Cheng}} \bibnamefont{and}
  \bibinfo{author}{\bibfnamefont{C.}~\bibnamefont{Xu}}, \bibinfo{journal}{Phys.
  Rev. B} \textbf{\bibinfo{volume}{94}}, \bibinfo{pages}{214415}
  (\bibinfo{year}{2016}),
  \urlprefix\url{https://link.aps.org/doi/10.1103/PhysRevB.94.214415}.

\bibitem[{\citenamefont{{Kachru} et~al.}(2017)\citenamefont{{Kachru},
  {Mulligan}, {Torroba}, and {Wang}}}]{Kachru2017_1}
\bibinfo{author}{\bibfnamefont{S.}~\bibnamefont{{Kachru}}},
  \bibinfo{author}{\bibfnamefont{M.}~\bibnamefont{{Mulligan}}},
  \bibinfo{author}{\bibfnamefont{G.}~\bibnamefont{{Torroba}}},
  \bibnamefont{and} \bibinfo{author}{\bibfnamefont{H.}~\bibnamefont{{Wang}}},
  \bibinfo{journal}{Physical Review Letters} \textbf{\bibinfo{volume}{118}},
  \bibinfo{eid}{011602} (\bibinfo{year}{2017}), \eprint{1609.02149}.

\bibitem[{\citenamefont{{Kachru} et~al.}(2016)\citenamefont{{Kachru},
  {Mulligan}, {Torroba}, and {Wang}}}]{Kachru2017_2}
\bibinfo{author}{\bibfnamefont{S.}~\bibnamefont{{Kachru}}},
  \bibinfo{author}{\bibfnamefont{M.}~\bibnamefont{{Mulligan}}},
  \bibinfo{author}{\bibfnamefont{G.}~\bibnamefont{{Torroba}}},
  \bibnamefont{and} \bibinfo{author}{\bibfnamefont{H.}~\bibnamefont{{Wang}}},
  \bibinfo{journal}{\prd} \textbf{\bibinfo{volume}{94}}, \bibinfo{eid}{085009}
  (\bibinfo{year}{2016}), \eprint{1608.05077}.

\bibitem[{\citenamefont{{Karch} et~al.}(2018)\citenamefont{{Karch}, {Tong}, and
  {Turner}}}]{Karch2018}
\bibinfo{author}{\bibfnamefont{A.}~\bibnamefont{{Karch}}},
  \bibinfo{author}{\bibfnamefont{D.}~\bibnamefont{{Tong}}}, \bibnamefont{and}
  \bibinfo{author}{\bibfnamefont{C.}~\bibnamefont{{Turner}}},
  \bibinfo{journal}{ArXiv e-prints}  (\bibinfo{year}{2018}),
  \eprint{1805.00941}.

\bibitem[{\citenamefont{{Fradkin} and {Kivelson}}(1996)}]{Fradkin1996}
\bibinfo{author}{\bibfnamefont{E.}~\bibnamefont{{Fradkin}}} \bibnamefont{and}
  \bibinfo{author}{\bibfnamefont{S.}~\bibnamefont{{Kivelson}}},
  \bibinfo{journal}{Nuclear Physics B} \textbf{\bibinfo{volume}{474}},
  \bibinfo{pages}{543} (\bibinfo{year}{1996}), \eprint{cond-mat/9603156}.

\bibitem[{\citenamefont{{Radicevic} et~al.}(2016)\citenamefont{{Radicevic},
  {Tong}, and {Turner}}}]{Radicevic2016}
\bibinfo{author}{\bibfnamefont{D.}~\bibnamefont{{Radicevic}}},
  \bibinfo{author}{\bibfnamefont{D.}~\bibnamefont{{Tong}}}, \bibnamefont{and}
  \bibinfo{author}{\bibfnamefont{C.}~\bibnamefont{{Turner}}},
  \bibinfo{journal}{Journal of High Energy Physics}
  \textbf{\bibinfo{volume}{12}}, \bibinfo{eid}{67} (\bibinfo{year}{2016}),
  \eprint{1608.04732}.

\bibitem[{\citenamefont{{Hui} et~al.}(2018)\citenamefont{{Hui}, {Mulligan}, and
  {Kim}}}]{Hui2018}
\bibinfo{author}{\bibfnamefont{A.}~\bibnamefont{{Hui}}},
  \bibinfo{author}{\bibfnamefont{M.}~\bibnamefont{{Mulligan}}},
  \bibnamefont{and} \bibinfo{author}{\bibfnamefont{E.-A.} \bibnamefont{{Kim}}},
  \bibinfo{journal}{\prb} \textbf{\bibinfo{volume}{97}}, \bibinfo{eid}{085112}
  (\bibinfo{year}{2018}), \eprint{1710.11137}.

\bibitem[{\citenamefont{{Wang} and {Senthil}}(2016)}]{WangSenthil2016}
\bibinfo{author}{\bibfnamefont{C.}~\bibnamefont{{Wang}}} \bibnamefont{and}
  \bibinfo{author}{\bibfnamefont{T.}~\bibnamefont{{Senthil}}},
  \bibinfo{journal}{\prb} \textbf{\bibinfo{volume}{94}}, \bibinfo{eid}{245107}
  (\bibinfo{year}{2016}), \eprint{1604.06807}.

\bibitem[{\citenamefont{{Sodemann} et~al.}(2017)\citenamefont{{Sodemann},
  {Kimchi}, {Wang}, and {Senthil}}}]{Sodemann2017}
\bibinfo{author}{\bibfnamefont{I.}~\bibnamefont{{Sodemann}}},
  \bibinfo{author}{\bibfnamefont{I.}~\bibnamefont{{Kimchi}}},
  \bibinfo{author}{\bibfnamefont{C.}~\bibnamefont{{Wang}}}, \bibnamefont{and}
  \bibinfo{author}{\bibfnamefont{T.}~\bibnamefont{{Senthil}}},
  \bibinfo{journal}{\prb} \textbf{\bibinfo{volume}{95}}, \bibinfo{eid}{085135}
  (\bibinfo{year}{2017}), \eprint{1609.08616}.

\bibitem[{\citenamefont{Qin et~al.}(2017)\citenamefont{Qin, He, You, Lu, Sen,
  Sandvik, Xu, and Meng}}]{dualnumerical}
\bibinfo{author}{\bibfnamefont{Y.~Q.} \bibnamefont{Qin}},
  \bibinfo{author}{\bibfnamefont{Y.-Y.} \bibnamefont{He}},
  \bibinfo{author}{\bibfnamefont{Y.-Z.} \bibnamefont{You}},
  \bibinfo{author}{\bibfnamefont{Z.-Y.} \bibnamefont{Lu}},
  \bibinfo{author}{\bibfnamefont{A.}~\bibnamefont{Sen}},
  \bibinfo{author}{\bibfnamefont{A.~W.} \bibnamefont{Sandvik}},
  \bibinfo{author}{\bibfnamefont{C.}~\bibnamefont{Xu}}, \bibnamefont{and}
  \bibinfo{author}{\bibfnamefont{Z.~Y.} \bibnamefont{Meng}},
  \bibinfo{journal}{Phys. Rev. X} \textbf{\bibinfo{volume}{7}},
  \bibinfo{pages}{031052} (\bibinfo{year}{2017}),
  \urlprefix\url{https://link.aps.org/doi/10.1103/PhysRevX.7.031052}.

\bibitem[{\citenamefont{{Witten}}(2003)}]{Witten2003}
\bibinfo{author}{\bibfnamefont{E.}~\bibnamefont{{Witten}}},
  \bibinfo{journal}{ArXiv High Energy Physics - Theory e-prints}
  (\bibinfo{year}{2003}), \eprint{hep-th/0307041}.

\bibitem[{\citenamefont{{Mross} et~al.}(2016)\citenamefont{{Mross}, {Alicea},
  and {Motrunich}}}]{Mross2016}
\bibinfo{author}{\bibfnamefont{D.~F.} \bibnamefont{{Mross}}},
  \bibinfo{author}{\bibfnamefont{J.}~\bibnamefont{{Alicea}}}, \bibnamefont{and}
  \bibinfo{author}{\bibfnamefont{O.~I.} \bibnamefont{{Motrunich}}},
  \bibinfo{journal}{Physical Review Letters} \textbf{\bibinfo{volume}{117}},
  \bibinfo{eid}{016802} (\bibinfo{year}{2016}), \eprint{1510.08455}.

\bibitem[{\citenamefont{{Mross} et~al.}(2017)\citenamefont{{Mross}, {Alicea},
  and {Motrunich}}}]{Mross2017}
\bibinfo{author}{\bibfnamefont{D.~F.} \bibnamefont{{Mross}}},
  \bibinfo{author}{\bibfnamefont{J.}~\bibnamefont{{Alicea}}}, \bibnamefont{and}
  \bibinfo{author}{\bibfnamefont{O.~I.} \bibnamefont{{Motrunich}}},
  \bibinfo{journal}{Physical Review X} \textbf{\bibinfo{volume}{7}},
  \bibinfo{eid}{041016} (\bibinfo{year}{2017}), \eprint{1705.01106}.

\bibitem[{\citenamefont{{Chen} et~al.}(2018)\citenamefont{{Chen}, {Son},
  {Wang}, and {Raghu}}}]{Chen2018}
\bibinfo{author}{\bibfnamefont{J.-Y.} \bibnamefont{{Chen}}},
  \bibinfo{author}{\bibfnamefont{J.~H.} \bibnamefont{{Son}}},
  \bibinfo{author}{\bibfnamefont{C.}~\bibnamefont{{Wang}}}, \bibnamefont{and}
  \bibinfo{author}{\bibfnamefont{S.}~\bibnamefont{{Raghu}}},
  \bibinfo{journal}{Physical Review Letters} \textbf{\bibinfo{volume}{120}},
  \bibinfo{eid}{016602} (\bibinfo{year}{2018}), \eprint{1705.05841}.

\bibitem[{\citenamefont{Karthik and Narayanan}(2017)}]{karthik2017}
\bibinfo{author}{\bibfnamefont{N.}~\bibnamefont{Karthik}} \bibnamefont{and}
  \bibinfo{author}{\bibfnamefont{R.}~\bibnamefont{Narayanan}},
  \bibinfo{journal}{Phys. Rev. D} \textbf{\bibinfo{volume}{96}},
  \bibinfo{pages}{054509} (\bibinfo{year}{2017}),
  \urlprefix\url{https://link.aps.org/doi/10.1103/PhysRevD.96.054509}.

\bibitem[{\citenamefont{{Banerjee} and {Radicevic}}(2014)}]{Banerjee2014}
\bibinfo{author}{\bibfnamefont{S.}~\bibnamefont{{Banerjee}}} \bibnamefont{and}
  \bibinfo{author}{\bibfnamefont{D.}~\bibnamefont{{Radicevic}}},
  \bibinfo{journal}{Journal of High Energy Physics}
  \textbf{\bibinfo{volume}{6}}, \bibinfo{eid}{168} (\bibinfo{year}{2014}),
  \eprint{1308.2077}.

\bibitem[{\citenamefont{{Jain} et~al.}(2013)\citenamefont{{Jain}, {Minwalla},
  and {Yokoyama}}}]{Jain2013}
\bibinfo{author}{\bibfnamefont{S.}~\bibnamefont{{Jain}}},
  \bibinfo{author}{\bibfnamefont{S.}~\bibnamefont{{Minwalla}}},
  \bibnamefont{and}
  \bibinfo{author}{\bibfnamefont{S.}~\bibnamefont{{Yokoyama}}},
  \bibinfo{journal}{Journal of High Energy Physics}
  \textbf{\bibinfo{volume}{11}}, \bibinfo{eid}{37} (\bibinfo{year}{2013}),
  \eprint{1305.7235}.

\bibitem[{\citenamefont{{Aharony} et~al.}(2013)\citenamefont{{Aharony},
  {Giombi}, {Gur-Ari}, {Maldacena}, and {Yacoby}}}]{Maldacena2013}
\bibinfo{author}{\bibfnamefont{O.}~\bibnamefont{{Aharony}}},
  \bibinfo{author}{\bibfnamefont{S.}~\bibnamefont{{Giombi}}},
  \bibinfo{author}{\bibfnamefont{G.}~\bibnamefont{{Gur-Ari}}},
  \bibinfo{author}{\bibfnamefont{J.}~\bibnamefont{{Maldacena}}},
  \bibnamefont{and} \bibinfo{author}{\bibfnamefont{R.}~\bibnamefont{{Yacoby}}},
  \bibinfo{journal}{Journal of High Energy Physics}
  \textbf{\bibinfo{volume}{3}}, \bibinfo{eid}{121} (\bibinfo{year}{2013}),
  \eprint{1211.4843}.

\bibitem[{\citenamefont{Naculich et~al.}(1990)\citenamefont{Naculich, Riggs,
  and Schnitzer}}]{Naculich1990}
\bibinfo{author}{\bibfnamefont{S.}~\bibnamefont{Naculich}},
  \bibinfo{author}{\bibfnamefont{H.}~\bibnamefont{Riggs}}, \bibnamefont{and}
  \bibinfo{author}{\bibfnamefont{H.}~\bibnamefont{Schnitzer}},
  \bibinfo{journal}{Physics Letters B} \textbf{\bibinfo{volume}{246}},
  \bibinfo{pages}{417 } (\bibinfo{year}{1990}), ISSN \bibinfo{issn}{0370-2693}.

\bibitem[{\citenamefont{Mlawer et~al.}(1991)\citenamefont{Mlawer, Naculich,
  Riggs, and Schnitzer}}]{Mlawer1991}
\bibinfo{author}{\bibfnamefont{E.~J.} \bibnamefont{Mlawer}},
  \bibinfo{author}{\bibfnamefont{S.~G.} \bibnamefont{Naculich}},
  \bibinfo{author}{\bibfnamefont{H.~A.} \bibnamefont{Riggs}}, \bibnamefont{and}
  \bibinfo{author}{\bibfnamefont{H.~J.} \bibnamefont{Schnitzer}},
  \bibinfo{journal}{Nuclear Physics B} \textbf{\bibinfo{volume}{352}},
  \bibinfo{pages}{863 } (\bibinfo{year}{1991}), ISSN \bibinfo{issn}{0550-3213}.

\bibitem[{\citenamefont{Naculich and Schnitzer}(2007)}]{Naculich2007}
\bibinfo{author}{\bibfnamefont{S.~G.} \bibnamefont{Naculich}} \bibnamefont{and}
  \bibinfo{author}{\bibfnamefont{H.~J.} \bibnamefont{Schnitzer}},
  \bibinfo{journal}{Journal of High Energy Physics}
  \textbf{\bibinfo{volume}{2007}}, \bibinfo{pages}{023} (\bibinfo{year}{2007}).

\bibitem[{\citenamefont{Wilson}(1974)}]{Wilson1974}
\bibinfo{author}{\bibfnamefont{K.~G.} \bibnamefont{Wilson}},
  \bibinfo{journal}{Phys. Rev. D} \textbf{\bibinfo{volume}{10}},
  \bibinfo{pages}{2445} (\bibinfo{year}{1974}).

\bibitem[{\citenamefont{Wilson}()}]{Wilson}
\bibinfo{author}{\bibfnamefont{K.~G.} \bibnamefont{Wilson}},
  \bibinfo{note}{{\it New Phenomena in Subnuclear Physics}, Springer US, 1977}.

\bibitem[{\citenamefont{Golterman et~al.}(1993)\citenamefont{Golterman, Jansen,
  and Kaplan}}]{Golterman1993}
\bibinfo{author}{\bibfnamefont{M.~F.} \bibnamefont{Golterman}},
  \bibinfo{author}{\bibfnamefont{K.}~\bibnamefont{Jansen}}, \bibnamefont{and}
  \bibinfo{author}{\bibfnamefont{D.~B.} \bibnamefont{Kaplan}},
  \bibinfo{journal}{Physics Letters B} \textbf{\bibinfo{volume}{301}},
  \bibinfo{pages}{219 } (\bibinfo{year}{1993}), ISSN \bibinfo{issn}{0370-2693}.

\bibitem[{\citenamefont{Coste and Lüscher}(1989)}]{Coste1989}
\bibinfo{author}{\bibfnamefont{A.}~\bibnamefont{Coste}} \bibnamefont{and}
  \bibinfo{author}{\bibfnamefont{M.}~\bibnamefont{Lüscher}},
  \bibinfo{journal}{Nuclear Physics B} \textbf{\bibinfo{volume}{323}},
  \bibinfo{pages}{631 } (\bibinfo{year}{1989}), ISSN \bibinfo{issn}{0550-3213}.

\bibitem[{\citenamefont{{Cordova}
  et~al.}(2017{\natexlab{c}})\citenamefont{{Cordova}, {Hsin}, and
  {Seiberg}}}]{Cordova2017}
\bibinfo{author}{\bibfnamefont{C.}~\bibnamefont{{Cordova}}},
  \bibinfo{author}{\bibfnamefont{P.-S.} \bibnamefont{{Hsin}}},
  \bibnamefont{and}
  \bibinfo{author}{\bibfnamefont{N.}~\bibnamefont{{Seiberg}}},
  \bibinfo{journal}{ArXiv e-prints}  (\bibinfo{year}{2017}{\natexlab{c}}),
  \eprint{1711.10008}.

\bibitem[{\citenamefont{{Chen} and {Zimet}}(2018)}]{ChenDual2018}
\bibinfo{author}{\bibfnamefont{J.-Y.} \bibnamefont{{Chen}}} \bibnamefont{and}
  \bibinfo{author}{\bibfnamefont{M.}~\bibnamefont{{Zimet}}},
  \bibinfo{journal}{ArXiv e-prints}  (\bibinfo{year}{2018}),
  \eprint{1806.04141}.

\end{thebibliography}

\onecolumngrid
\appendix
\section{Integration over the \boldmath $\SO(3)$ gauge field}
\label{App:IntegrationSO(3)}
To integrate out the $\SO(3)$ gauge field in the action Eq.
\ref{Eq:Lattice_Boson_Fermion_theory}, we only need to focus on
the $S_{\rm bg}$ and $S_{\rm fg}$ parts of the action:
\begin{align}
& \int D[O^{n\mu}] ~ e^{-S_{\rm bg}[v, O]} e^{ - S_{\rm fg}[\chi, O] }
\nonumber \\
& = \exp \left(-\sum_i \sum_n M \bar{\chi}_{n,i} \chi_{n,i}\right)
\nonumber \\
& ~~~~ \times \prod_n \prod_{\mu=x,y,z} \left(\sum_{l=0}^\infty
\sum_{m=0}^3 \frac{J^l}{l!} \frac{(-1)^m}{m!} \int_{\SO(3)}
dO^{n\mu} \left(v_{n+\mu,i} O^{n\mu}_{ij} v_{n,j} \right)^l \left(
\bar{\chi}_{n+\mu,i} (\sigma^\mu -R) O^{n\mu}_{ij} \chi_{n,j}
\right)^m
 \right),
\label{Eq:OIntegration_App}
\end{align}
Generically, since the fermion field $\chi_n$ on each site carries a two-fold spinor index and a three-fold $\SO(3)$ color index, the terms in the expansion are expected to vanish because of Fermi statistics only when $m>6$. However, when we take $R=-1$ (as we did in the main text), all the terms with $m>3$ will vanish. The simplification comes from the fact that $\sigma^\mu-R$ for any fixed $\mu$ is a rank-1 matrix acting on the spinor space of the fermion $\chi_n$ when $R=-1$. In another word, for a specific link labeled by $n$ and $\mu$, $(\sigma^\mu-R)$ only ``hops" a single spinor mode of $\chi_{n,i}$ out of its two dimensional spinor space from site $n$ to $n+\mu$. Therefore, the terms with $m>3$ all vanish because of Fermi statistics in Eq. \ref{Eq:OIntegration_App}. Since the one spinor mode that is singled out by $\sigma^\mu-R$ for is different in different directions $\mu$, the fermion kinetic term $\sigma^\mu-R$ overall {\it does not} leave any fermion mode non-propagating. We will choose $R=-1$ throughout the discussion.

In the following, we will tackle the
integration in Eq. \ref{Eq:OIntegration_App} for the terms with $m=0,1,2,3$ separately. The
following identity on the integration over the $\SO(3)$ group will
be helpful:
\begin{align}
&  \int_{\SO(3)} dO ~ (y^\T Ox)^n (v^\T Ou) (s^\T Or)
\nonumber \\
& =
\begin{cases}
      \frac{1}{2 (n+2)} \left[ x\cdot (u \times r) \right] \left[ y \cdot (v \times s) \right]
      & n ~~ \rm{odd}
\\
\frac{1}{2(n+1)(n+3)} \left\{
(n+2) (u^\T r) (v^\T s)
- n \left[
(x^\T u)  (x^\T r) (v^\T s) + (y^\T v)  (y^\T s) (u^\T r)
\right]
+3n (x^\T u)  (x^\T r) (y^\T v)  (y^\T s)
\right\}
       & n ~~ \rm{even}
\end{cases}.
\label{Eq:SO(3)Ave_XY_UV_RS}
\end{align}
Here, the $\SO(3)$ matrix $O$ is being integrated under the Haar
measure of $\SO(3)$. $x,y\in\R^3$ are 3-component vectors of unit
length, while $r,s,v,u\in\R^3$ are any 3-component vectors of
arbitrary length.

For the terms with $m=0$ in Eq. \ref{Eq:OIntegration_App}, we have
\begin{align}
\sum_{l=0}^\infty  \frac{J^l}{l!}  \int_{\SO(3)} dO^{n\mu}
\left(v_{n+\mu}^\T O^{n\mu} v_{n} \right)^l = \sum_{l~ {\rm even}}
\frac{J^l}{l!} \frac{1}{l+1} =\frac{\sinh J}{J}.
\label{Eq:Integrate_SO3_Gauge_Field_m0}
\end{align}
For $m=1$, we have
\begin{align}
& ~~~~ \sum_{l=0}^\infty  - \frac{J^l}{l!}  \int_{\SO(3)}
dO^{n\mu} \left(v_{n+\mu}^\T O^{n\mu} v_{n} \right)^l \left(
\bar{\chi}_{n+\mu,i} (\sigma^\mu -R) O^{n\mu}_{ij} \chi_{n,j}
\right)
\nonumber \\
& = \sum_{l~ {\rm odd}} - \frac{J^l}{l!} \frac{1}{l+2} \left(
(\bar{\chi}_{n+\mu,i} v_{n+\mu,i}) (\sigma^\mu -R) (\chi_{n,j}
v_{n,j}) \right)
\nonumber \\
& = - \frac{J \cosh J -\sinh J}{J^2} (\bar{\chi}_{n+\mu,i}
v_{n+\mu,i}) (\sigma^\mu -R) (\chi_{n,j} v_{n,j}) .
\label{Eq:Integrate_SO3_Gauge_Field_m1}
\end{align}
For $m=2$, we have
\begin{align}
& ~~~~ \sum_{l=0}^\infty   \frac{J^l}{l!} \frac{1}{2}
\int_{\SO(3)} dO^{n\mu} \left(v_{n+\mu}^\T O^{n\mu} v_{n}
\right)^l \left( \bar{\chi}_{n+\mu,i} (\sigma^\mu -R)
O^{n\mu}_{ij} \chi_{n,j} \right)^2
\nonumber \\
& =  \sum_{l ~ {\rm odd }}   \frac{J^l}{l!} \frac{1}{2} \frac{1}{2(l+2)}
\varepsilon_{ii'i''}\varepsilon_{jj'j''}
\left( \bar{\chi}_{n+\mu,i} (\sigma^\mu -R) \chi_{n,j} \right)
\left( \bar{\chi}_{n+\mu,i'} (\sigma^\mu -R) \chi_{n,j'} \right)
v_{n+\mu,i''} v_{n, j''}
\nonumber \\
& =  \frac{J \cosh J -\sinh J}{4J^2}
\varepsilon_{ii'i''}\varepsilon_{jj'j''}
\left( \bar{\chi}_{n+\mu,i} (\sigma^\mu -R) \chi_{n,j} \right)
\left( \bar{\chi}_{n+\mu,i'} (\sigma^\mu -R) \chi_{n,j'} \right)
v_{n+\mu,i''} v_{n, j''}.
\nonumber \\
& =  \frac{J \cosh J -\sinh J}{4J^2}
\varepsilon_{ii'i''}\varepsilon_{jj'j''}
\J^{n\mu}_{ij}\J^{n\mu}_{i'j'}
v_{n+\mu,i''} v_{n, j''}.
\label{Eq:Integrate_SO3_Gauge_Field_m2}
\end{align}
Notice that the even $l$ contributions in the equation above all
vanish because of fermion statistics. Lastly, for $m=3$, we have
\begin{align}
& ~~~~ \sum_{l=0}^\infty  - \frac{J^l}{l!} \frac{1}{6}
\int_{\SO(3)} dO^{n\mu} \left(v_{n+\mu}^\T O^{n\mu} v_{n}
\right)^l \left( \bar{\chi}_{n+\mu,i} (\sigma^\mu -R)
O^{n\mu}_{ij} \chi_{n,j} \right)^3
\nonumber \\
& = \sum_{l=0}^\infty  - \frac{J^l}{l!} \frac{1}{6} \int_{\SO(3)}
dO^{n\mu} \left(v_{n+\mu}^\T O^{n\mu} v_{n} \right)^l \left(
\bar{\chi}_{n+\mu,i} (\sigma^\mu -R) \chi_{n,i} \right)^3
\nonumber \\
& = - \frac{1}{6} \frac{\sinh J}{J}
\left( \bar{\chi}_{n+\mu,i} (\sigma^\mu -R) \chi_{n,i} \right)^3,
\nonumber \\
& = - \frac{1}{36} \frac{\sinh J}{J}
\varepsilon_{ii'i''}\varepsilon_{jj'j''}
\J^{n\mu}_{ij} \J^{n\mu}_{i'j'} \J^{n\mu}_{i''j''}
\label{Eq:Integrate_SO3_Gauge_Field_m3}
\end{align}
where we've used the fact that $\left( \bar{\chi}_{n+\mu,i}
(\sigma^\mu -R) O^{n\mu}_{ij} \chi_{n,j} \right)^3 = (\det
O^{n\mu}) \left( \bar{\chi}_{n+\mu,i} (\sigma^\mu -R) \chi_{n,i}
\right)^3$ for $R=-1$ and $\det O^{n\mu} =1$. Now, we can conclude that
\begin{align}
& \int D[O^{n\mu}] ~ e^{-S_{\rm bg}[v, O]} e^{ - S_{\rm fg}[\chi,
O] }
\nonumber \\
&  = \exp \left( - \sum_n M \bar{\chi}_{n,i} \chi_{n,i}\right)
\times \prod_{n,\mu}\left[ \frac{\sinh J}{J}  - \frac{J \cosh J
-\sinh J}{J^2} (\bar{\chi}_{n+\mu,i} v_{n+\mu,i}) (\sigma^\mu -R)
(\chi_{n,j} v_{n,j})   \right. \nonumber
  \\
& \left. +  \frac{J \cosh J -\sinh J}{4J^2}
\varepsilon_{ii'i''}\varepsilon_{jj'j''} \J^{n\mu}_{ij}
\J^{n\mu}_{i'j'} v_{n+\mu,i''} v_{n, j''}  - \frac{1}{36}
\frac{\sinh J}{J} \varepsilon_{ii'i''}\varepsilon_{jj'j''}
\J^{n\mu}_{ij} \J^{n\mu}_{i'j'} \J^{n\mu}_{i''j''}  \right]
\nonumber
\\
& = e^{-S_{\rm bfg}[v,\chi]} \times \prod_{n,\mu} \frac{\sinh J}{J}
\end{align}
with the effective action $S_{\rm bfg}[v,\chi]$:
\begin{align}
S_{\rm bfg}[v,\chi] = & \sum_n \left\{ M \bar{\chi}_{n,i}
\chi_{n,i} + K (\bar{\chi}_{n+\mu,i} v_{n+\mu,i}) (\sigma^\mu -R)
(\chi_{n,j} v_{n,j}) - \frac{K}{4}
\varepsilon_{ii'i''}\varepsilon_{jj'j''} \J^{n\mu}_{ij}
\J^{n\mu}_{i'j'} v_{n+\mu,i''} v_{n, j''} \right.
\nonumber \\
& \left.
+ \frac{1 - K^2}{36}
\varepsilon_{ii'i''}\varepsilon_{jj'j''}
\J^{n\mu}_{ij} \J^{n\mu}_{i'j'} \J^{n\mu}_{i''j''}
\right\}.
\end{align}
where we've defined the variable $K = \frac{J \cosh J - \sinh J}{J
\sinh J}$. In obtaining this effective action, we've applied the
identity $(\bar{\chi}_{n+\mu,i} v_{n+\mu,i}) (\sigma^\mu -R)
(\chi_{n,j} v_{n,j}) \times \left(
\varepsilon_{ii'i''}\varepsilon_{jj'j''} \J^{n\mu}_{ij}
\J^{n\mu}_{i'j'} v_{n+\mu,i''} v_{n, j''} \right) = \frac{1}{9}
\varepsilon_{ii'i''}\varepsilon_{jj'j''} \J^{n\mu}_{ij}
\J^{n\mu}_{i'j'} \J^{n\mu}_{i''j''} $ for $R=-1$.

\section{Derivation of the alternative dual model}
\label{App:Derivation_Dual}
Without the cancellation condition Eq. \ref{Eq:Cancellation_Condition}, we can generally write $e^{-S_{\rm bfg}[v,\chi] - S_{\rm int}[v,\chi]}$ as
\begin{align}
e^{-S_{\rm bfg}[v,\chi] - S_{\rm int}[v,\chi]}= & \exp\left(-\sum_{n}  M \bar{\chi}_{n,i} \chi_{n,i} \right) \times  
 \prod_{n,\mu} \Big\{1 - K (\bar{\chi}_{n+\mu,i} v_{n+\mu,i}) (\sigma^\mu -R) (\chi_{n,j} v_{n,j}) 
\nonumber \\
&
\left.
 + \frac{K-U_1}{4}\varepsilon_{ii'i''}\varepsilon_{jj'j''} \J^{n\mu}_{ij} \J^{n\mu}_{i'j'} 
v_{n+\mu,i''} v_{n, j''} 
 - \frac{1+U_2- U_1 K}{36}
\varepsilon_{ii'i''}\varepsilon_{jj'j''}
\J^{n\mu}_{ij} \J^{n\mu}_{i'j'} \J^{n\mu}_{i''j''} 
\right\},
\end{align}
We've introduced the fermion field $\xi_n$ in Eq. \ref{Eq:Fermion_Rewriting} and fermion field $\eta_{n,a}$ (with $a=1,2$) in Eq. \ref{Eq:eta_fermion_Def}. These three fermion fields are orthogonal to each other in the $\SO(3)$ color space. Under a fixed configuration of the vector boson $v_n$, we can write
\begin{align}
& e^{- S_{\rm bfg} - S_{\rm int}} 
\nonumber \\
& =   \exp \left[ \sum_{n}  \left( M' \bar{\xi}_{n,i} \xi_{n,i} + M'' \bar{\eta}_{n,1} \eta_{n,1} + M'' \bar{\eta}_{n,2} \eta_{n,2} \right)  \right] \times 
 \prod_{n,\mu} \Big\{ 1 - \bar{\xi}_{n+\mu}  (\sigma^\mu -R) \xi_{n} 
 \nonumber \\
& ~~~
 + \left( \bar{\eta}_{n+\mu,1}  (\sigma^\mu -R) \eta_{n,1} \right) \left( \bar{\eta}_{n+\mu,2}  (\sigma^\mu -R) \eta_{n,2} \right) 
 - \frac{1+U_2-U_1K}{(K-U_1)K}
 \left( \bar{\xi}_{n+\mu}  (\sigma^\mu -R) \xi_{n} \right)
  \left( \bar{\eta}_{n+\mu,1}  (\sigma^\mu -R) \eta_{n,1} \right) 
  \left( \bar{\eta}_{n+\mu,2}  (\sigma^\mu -R) \eta_{n,2} \right) \Big\}
\label{Eq:action_XiEta}
\end{align}
where 
\begin{align}
M' = M/K, ~~~ M'' = M (K- U_1)^{-\frac{1}{2}}.
\end{align}
Notice that we've been assuming that $K>U_1$ in this discussion. To simplify the expression further, a trick is to introduce an $\SO(2)$ gauge field $U^{n\mu}$, represented by a $2\times2$ orthogonal matrix $U^{n\mu}_{ab}$ with $a,b=1,2$, on every link:
\begin{align}
& e^{- S_{\rm bfg} - S_{\rm int}} = 
\exp \left[ - \sum_{n}  \left( M' \bar{\xi}_{n} \xi_{n} + M'' \bar{\eta}_{n,a} \eta_{n,a} \right)  \right] 
\times 
 \prod_{n,\mu} \Big\{ \int dU^{n\mu} \exp\left(- \bar{\xi}_{n+\mu}  (\sigma^\mu -R) \xi_{n}  \right)
\nonumber 
\\
& \times \exp\Big(- \bar{\eta}_{n+\mu,a}  (\sigma^\mu -R) U^{n\mu}_{ab} \eta_{n,b}  \Big) \times \exp \Big( 
-V 
\left( \bar{\xi}_{n+\mu}  (\sigma^\mu -R) \xi_{n} \right)
  \left( \bar{\eta}_{n+\mu,1}  (\sigma^\mu -R) \eta_{n,1} \right) 
  \left( \bar{\eta}_{n+\mu,2}  (\sigma^\mu -R) \eta_{n,2} \right) 
\Big)
\Big\}
\label{Eq:action_XiEtaU}
\end{align}
with $V= \frac{1+U_2  -K^2}{(K-U_1)K}$. Here, the repeated $\SO(2)$ indices $a,b$ are automatically summed from 1 to 2.
Although this rewriting is introduced as a trick, its physical meaning is discussed in Sec. \ref{Sec:Alternative_Dual}. To show that Eq. \ref{Eq:action_XiEtaU} is equivalent to Eq. \ref{Eq:action_XiEta}, we simply need to perform a Taylor expansion to the exponential term $\exp\Big(- \bar{\eta}_{n+\mu,a}  (\sigma^\mu -R) U^{n\mu}_{ab} \eta_{n,b}  \Big)$ in Eq. \ref{Eq:action_XiEtaU}. Here, we've also chosen $R=-1$ which in this case leads to only 3 non-vanishing terms, i.e the zeroth, the first and the second order terms, in the expansion. Other terms vanish because of Fermi statistics. Now, we can perform the integration $\int dU^{n\mu}$ term by term. The first order term further vanishes because it contains an odd power of $U^{n\mu}$. The integration for the zeroth and the second order terms are also simple because they are in fact independent of $U^{n\mu}$. Putting these together, we can verify that Eq. \ref{Eq:action_XiEtaU} is consistent with Eq. \ref{Eq:action_XiEta}. Eq. \ref{Eq:action_XiEtaU} is essentially Eq. \ref{Eq:Alternative_Dual} after regrouping different terms on the right hand side of the equation.

\section{Lattice duality with $Z_2$ gauge field}\label{sec:Z2 gauge}

In this appendix, we provide the lattice construction of a similar boson-fermion duality with the gauge group (on the boson side) reduced from $\O(3)$ to $Z_2$. Loosely speaking, we will show that a single-component real scalar boson coupled to an ``$\O(1)_1$ Chern-Simons'' gauge field is dual to a free Majorana fermion in 2+1D. The precise meaning of the ``$\O(1)_1$ Chern-Simons'' term is actually a $Z_2$ gauge field coupled to massive Majorana fermions in a Chern band with Chern number $C=1$, which is also known as the Ising topological order.

We start with the path integral formulation on the 3D Euclidian spacetime lattice. On the bosonic side of the duality, we introduce on each site $n$ a single-component scalar boson $\sigma_n$ (which can treated as a $Z_2$ variable $\sigma_n=\pm1$) and two real Grassmanian variables $\chi_{n}=(\chi_{n1},\chi_{n2})^\T$. They both couple to a $Z_2$ gauge field $B^{n\mu}=\pm1$ on the link $n\mu$ that connects site $n$ and $n+ \mu$ (where $\mu=0,1,2$ labels the link direction). The partition function of the Euclidean lattice path integral is given by
\begin{equation}\label{eq: Z2 Z_b}
Z_b=\int D[\chi]\sum_{[\sigma,B]}e^{-S_\text{bg}[\sigma,B]-S_\text{fg}[\chi,B]},
\end{equation}
where the actions $S_\text{bg}$ and $S_\text{fg}$ are given by
\begin{equation}
\begin{split}
S_\text{bg}[\sigma,B]&=-J\sum_{n\mu}\sigma_{n+ \mu}B^{n\mu}\sigma_{n},\\
S_\text{fg}[\chi,B]&=\sum_{n\mu}\bar{\chi}_{n+ \mu}(\gamma^\mu-R)B^{n\mu}\chi_{n}+\sum_{n}M\bar{\chi}_n\chi_n.
\end{split}
\end{equation}
Here the gamma matrices are defined as $(\gamma^0,\gamma^1,\gamma^2)=(\sigma^2,\sigma^3,\sigma^1)$ and $\bar{\chi}_n\equiv\chi_n^\T\gamma^0$. The  Majorana Chern number of the lattice fermion $\chi$ is still given by Eq.~\ref{Eq:Majorana_Chern_Number}. The theory describes the an Ising Higgs model twisted by auxiliary Majorana fermions. The interaction among auxiliary fermions $\chi$ can be circumvented when the gauge group is $Z_2$, which simplifies the derivation of the duality, similar to the case of $\O(3)$ gauge group in Eq.~\ref{Eq:Lattice_Boson_Fermion_theory_O3}. The partition function $Z_b$ in Eq.~\ref{eq: Z2 Z_b} can be expanded on the lattice to the following form
\begin{equation}\label{eq: Z TWV}
\begin{split}
&Z_b=\int D[\chi]\sum_{[\sigma,B]}\prod_{n\mu}T_{n\mu}(R)[\chi,B]W_{n\mu}(J)[\sigma,B]\prod_{n}V_n(M)[\chi],\\
&T_{n\mu}(R)[\chi,B]=e^{-\bar{\chi}_{n+ \mu}(\gamma^\mu-R)B^{n\mu}\chi_{n}}=1-\bar{\chi}_{n+ \mu}(\gamma^\mu-R)B^{n\mu}\chi_{n},\\
&W_{n\mu}(J)[\sigma,B]=e^{J\sigma_{n+ \mu}B^{n\mu}\sigma_{n}}\propto 1+(\tanh J)\sigma_{n+ \mu}B^{n\mu}\sigma_{n},\\
&V_{n}(M)[\chi]=e^{-M\bar{\chi}_n\chi_n}.
\end{split}
\end{equation}
In the expansion of $T_{n\mu}(R)[\chi,B]$, we have assumed  $R=-1$ such that the expansion terminates at the quadratic order. On each link $n\mu$, we can first integrating out the $Z_2$ gauge field $B^{n\mu}=\pm1$, and arrive at a new link term
\begin{equation}\label{eq: T'_chi}
T'_{n\mu}(R,J)[\chi,\sigma]=\sum_{B^{n\mu}}T_{n\mu}(R)[\chi,B]W_{n\mu}(J)[\sigma,B]=1-(\tanh J)\;\sigma_{n+ \mu}\bar{\chi}_{n+ \mu}(\gamma^\mu-R)\chi_{n}\sigma_{n}.
\end{equation}
With this, the partition function in Eq.~\ref{eq: Z TWV} reduces to
\begin{equation}\label{eq: Z T'V}
Z_b=\int D[\chi]\sum_{[\sigma]}\prod_{n\mu}T'_{n\mu}(R,J)[\chi,\sigma]\prod_{n}V_{n}(M)[\chi].
\end{equation}
Integrating out the scalar (Ising) field $\sigma_n$ simply imposes the current conservation of the fermion $\chi_n$, which is already built-in in the fermion path integral formalism. If we redefine a new set of real Grassmannian variables
\begin{equation}
\xi_{n}=\sqrt{\tanh J}\;\chi_n\sigma_n,
\end{equation}
the partition function in Eq.~\ref{eq: Z T'V} will simply become a theory of $\xi_n$ fermion with renormalized mass,
\begin{equation}
\begin{split}
&Z_f=\int D[\xi]\prod_{n\mu}T_{n\mu}(R)[\xi]\prod_{n}V_{n}(M')[\xi],\\
&T_{n\mu}(R)[\xi]=1-\bar{\xi}_{n+ \mu}(\gamma^\mu-R)\xi_{n}=e^{-\bar{\xi}_{n+ \mu}(\gamma^\mu-R)\xi_{n}},\\
&V_{n}(M')[\xi]=e^{-M'\bar{\xi}_n\xi_n}.
\end{split}
\end{equation}
where the renormalized mass $M'$ is given by
\begin{equation}
M'=M/\tanh J.
\end{equation}
Now we have arrived at a theory that exactly describes a free Majorana on the Euclidean spacetime lattice, which can be equivalently written in the action form as follows
\begin{equation}\label{eq: Z2 Z_f}
\begin{split}
&Z_f=\int D[\xi]e^{-S_\text{f}[\xi]},\\
&S_\text{f}[\xi]=\sum_{n\mu}\bar{\xi}_{n+ \mu}(\gamma^\mu-R)\xi_{n}+\sum_{n}M'\bar{\xi}_n\xi_n.
\end{split}
\end{equation}
Thus we have established an exact lattice duality between the twisted Ising Higgs model $Z_b$ in Eq.~\ref{eq: Z2 Z_b} and the free Majorana model $Z_f$ in Eq.~\ref{eq: Z2 Z_f}. This exact mapping is valid regardless whether the model $Z_b$ (or its dual model $Z_f$) is at the critical point or not. Knowing that the dual Majorana fermion $\xi$ has a critical point at $R=-1$ and $M'=3$, the original model $Z_b$ should have the same critical point at $R=-1$, $M=M' \tanh J=3 \tanh J$. For any positive coupling $J$, $M$ is always smaller than 3. If we choose $J$ such that $1<M<3$, as we integrate out the fermion $\chi$, the model $Z_b$ can be interpreted as a $Z_2$ matter field coupled to a $Z_2$ gauge field (with Ising topological order).

\section{Duality towards a free Majorana fermion plus a decoupled $Z_2$ gauge field}
\label{App:Majorana_Z2}
In Ref. \onlinecite{Cordova2017}, an IR duality between the critical vector boson coupled to an $\O(3)^{0}_{1,1}$ Chern-Simons gauge theory (following the notation of Ref. \onlinecite{Cordova2017}) and the theory with a free Majorana fermion plus a decoupled $Z_2$ gauge theory is proposed. Before we discuss a related lattice duality, a brief explanation of the $\O(3)^{0}_{1,1}$ Chern-Simons term is in order. In the convention of Ref. \onlinecite{Cordova2017}, $\O(3)^{0}_{1,1}$ Chern-Simons term is an $\O(3)$ Chern-Simons term at level-1 plus another topological term $f[w_1]$ at level-1. Here, $w_1$ is the 1st Stiefel-Whitney class of the $\O(3)$ gauge bundle, which in this case can be identified with the $Z_2$ gauge fields obtained from restricting the $\O(3) = \SO(3) \times Z_2$ gauge group to its $Z_2$ subgroup. When we viewed $w_1$ as a $Z_2$ gauge field, $f[w_1]$ is the topological term generated by coupling the $Z_2$ gauge field $w_1$ to a single copy of $Z_2$-charged massive Majorana fermion in a band structure with Chern number 1 (or equivalently to a single copy of $p+ip$ superconductor with the fermions carrying a $Z_2$ charge). Since we can decompose the gauge group according to $\O(3)= \SO(3) \times Z_2$, we can rewrite following Ref. \onlinecite{Cordova2017} that 
\begin{align}
\O(3)^{0}_{1,1}~\text{Chern-Simons term} = \SO(3)_{1}~\text{Chern-Simons term}+ \pi w_2[\SO(3)] \cup w_1+ 3 f[w_1],
\end{align}
where $w_1$ is identified with the $Z_2$ subgroup of the $\O(3)$ gauge group and $w_2[\SO(3)]$ is the 2nd Stiefel-Whitney class of the $\SO(3)$ part of the gauge field. From this relation, one can see that the $\O(3)^{0}_{1,1}$ Chern-Simons term can be generated by coupling 3 copies of massive Majorana fermions (forming an vector representation under the $\O(3)$ gauge group) each in a band structure with Chern number 1 to an $\O(3)$ gauge field. As we will see, this understanding of the $\O(3)^{0}_{1,1}$ Chern-Simons term will also be helpful in understanding the IR nature of the lattice model we will discuss in the following.

This proposed IR duality in Ref. \onlinecite{Cordova2017} is similar to the lattice duality studied in Sec. \ref{Sec:O(3)Duality} on the corresponding ``boson sides" as they both describe a vector boson coupled to an $\O(3)$ Chern-Simons gauge theory. However, on the ``fermion side", while both studies contain the theory of a free Majorana fermion, the proposed duality of Ref. \onlinecite{Cordova2017} also includes an extra decoupled $Z_2$ gauge theory. Inspired by the proposed IR duality, we will introduce a lattice model that is slightly different from Eq. \ref{Eq:Lattice_Boson_Fermion_theory_O3} and construct an exact mapping to a dual theory containing a free Majorana fermion and a decoupled $Z_2$ gauge theory. We will agree that this lattice duality is connected to the IR duality between the critical vector boson coupled to an $\O(3)^{0}_{1,1}$ Chern-Simons gauge theory and a free Majorana fermion plus a decoupled $Z_2$ gauge theory.

Now, we introduce the ingredients of the lattice model. We consider a model with the same degrees of freedom as the model Eq. \ref{Eq:Lattice_Boson_Fermion_theory_O3}: a vector boson field $v_n$ and a Majorana fermion field $\chi_n$ on each site both coupled to the $\O(3)$ gauge field $O^{n\mu}$ on the links. Since $\O(3) =\SO(3) \times Z_2$, we can always decompose the O(3) gauge field as $O^{n\mu} = B^{n\mu} \tilde{O}^{n\mu}$, where $B^{n\mu}=\pm 1$ describes a $Z_2$ gauge field and $\tilde{O}^{n\mu}$ is an $\SO(3)$ matrix that describes a $\SO(3)$ gauge field on the link $n\mu$. The integration over the $\O(3)$ gauge field $O^{n\mu}$ is equivalent to the integration over the $\SO(3)$ gauge field $\tO^{n\mu}$ under the Haar measure of $\SO(3)$ followed by the summation over $B^{n\mu}=\pm 1$, i.e. $\int D[O^{n\mu}] = \int D[\tO^{n\mu}]\sum_{[B^{n\mu}]}$, where $\sum_{[B^{n\mu}]}$ represents the summation over all configuration of $B^{n\mu}=\pm 1$. We will include the actions $S_{\rm bg}'$ and $S_{\rm fg}'$ introduced in Sec. \ref{Sec:O(3)Duality} in the current lattice model. Using these new gauge field variables, we can rewrite them as 
\begin{align}
S_{\rm bg}'[v,O] = S_{\rm bg}'[v,\tO,B] = \sum_n \sum_{\mu=x,y,z} -J~ v_{n+\mu,i} B^{n\mu} \tO_{ij}^{n\mu} v_{n,j}
\end{align}
and 
\begin{align}
S'_{\rm fg}[\chi,O] = S_{\rm fg}'[\chi,\tO,B] = 
 \sum_n  \sum_{\mu = x,y,z } \bar{\chi}_{n+\mu,i} (\sigma^\mu -R) B^{n\mu} \tO^{n\mu}_{ij} \chi_{n,j}
+  M \sum_n \bar{\chi}_{n,i}  \chi_{n,i}.
\end{align}
While the model studied in Sec. \ref{Sec:O(3)Duality} only contains the terms 
$S_{\rm bg}'$ and $S_{\rm fg}'$, we will further introduce an interaction term 
$S_{\rm int}' $ and a gauge field term $S_{Z_2}'$ for the current model of interest. The interaction terms $S_{\rm int}' $ is given by:
\begin{align}
& S_{\rm int}'[\chi, v, B] =  \frac{U_1}{4} \sum_{n,\mu}
\varepsilon_{ii'i''}\varepsilon_{jj'j''} B^{n\mu}\J^{n\mu}_{ij} \J^{n\mu}_{i'j'}
v_{n+\mu,i''} v_{n, j''}
+ \frac{U_2}{36}
\sum_{n,\mu}
\varepsilon_{ii'i''}\varepsilon_{jj'j''}
B^{n\mu} \J^{n\mu}_{ij} \J^{n\mu}_{i'j'} \J^{n\mu}_{i''j''}.
\end{align}
Notice that even though $S_{\rm int}'[\chi, v, B] $ only depends on the $Z_2$ gauge field $B^{n\mu}$, it is fully $\O(3)$ gauge-invariant. In fact, one can view $S_{\rm int}'$ as the ``$\O(3)$-gauged" version of Eq. \ref{Eq:S_SO3_int_def}. The $Z_2$ gauge field $B^{n\mu}$ can have its only dynamics described by the standard $Z_2$ lattice gauge theory action:
\begin{align}
S_{Z_2}'[B] = \sum_{{\rm plaquette}~p} ~t \prod_{{\rm link}~l \in p} B^l,
\label{Eq:Z2_Gauge_Action}
\end{align}
where $p$ labels the 2-dimensional square plaquettes of the 3D spacetime cubic lattice, the product $\prod_{{\rm link}~l \in p}$ represents the product over all links $l$  that belong to the plaquette $p$, and $t$ denotes the coupling constant of the $Z_2$ gauge field $B^{n\mu}$. Notice that such a pure gauge dynamical term was not introduced in any previous lattice models studied in the main text where the only gauge field dynamics were thought of as generated by the coupling to the Majorana fermions. In the discussion here, we not only couple the gauge field to the fermions $\chi_n$, but also include the term $S_{Z_2}'[B]$ into consideration. As we will see, such a gauge dynamical term will help keeping the $Z_2$ gauge field deconfined across the duality.
Having introduced all the ingredients, we can write down the model of interest
\begin{align}
Z_b'' = \sum_{[B^{n\mu}]} \int D[\tO^{n\mu}] \int D[\chi_{n,i}] \int D[v_n] e^{-S'_{\rm
bg} - S'_{\rm fg}-S_{\rm int}'-S_{Z_2}'}. \label{Eq:Lattice_Boson_Fermion_theory_O3_Z2}
\end{align} 

To construct the lattice duality, we first integrate out the $\SO(3)$ gauge field $\tO_{n\mu}$ in Eq. \ref{Eq:Lattice_Boson_Fermion_theory_O3_Z2}. The technical details of this integration very much follow those in Sec. \ref{Sec:SO(3)Duality} and Appendix \ref{App:IntegrationSO(3)}. Similar to Sec. \ref{Sec:SO(3)Duality}, by choosing the parameters Eq. \ref{Eq:Cancellation_Condition}, we have 
\begin{align}
& \int D[\tO^{n\mu}] e^{-S'_{\rm bg} - S'_{\rm fg} - S'_{\rm int }}
 =  \exp\left(-\sum_{n}  M \bar{\chi}_{n,i} \chi_{n,i} \right) \times  
 \prod_{n,\mu} \Big[ 1 - K (\bar{\chi}_{n+\mu,i} v_{n+\mu,i}) (\sigma^\mu -R) (\chi_{n,j} v_{n,j})  \Big].
\end{align}
Interestingly, even though all of $S'_{\rm bg}$, $S'_{\rm fg}$ and $S'_{\rm int }$ depend on the $Z_2$ gauge field $B^{n\mu}$, the right hand side of this equation is independent of $B^{n\mu}$ as the result of both the integration over the $\SO(3)$ gauge field $\tO^{n\mu}$ and the parameter choice Eq. \ref{Eq:Cancellation_Condition}. Again, we introduce the Majorana fermion field $\xi_n$ following Eq. \ref{Eq:Fermion_Rewriting} and integrate out the Majorana fermion fields $\xi'_n$ and $\xi''_n$ that are orthogonal to $\xi_n'$ in the $\O(3)$ color space. We obtain that
\begin{align}
&  \sum_{[B^{n\mu}]}  \int D[\tO^{n\mu}] \int D[\chi_{n,i}] e^{-S_{\rm bg}' - S_{\rm fg}' - S_{\rm int }'-S'_{Z_2}}
\nonumber \\
& = \mathcal{N}'' \sum_{[B^{n\mu}]}  \int D[\xi_n]
\exp\left(-\sum_{n,\mu} \bar{\xi}_{n+\mu}  (\sigma^\mu -R) \xi_n  - \sum_n M' \bar{\xi}_{n} \xi_{n}\right) \times  \exp(-S'_{Z_2}),
\label{Eq:Dual_Fermion_O3_Z2}
\end{align}
where $M' = M/K$ and $\mathcal{N}''$ is an overall normalization constant. We notice that the right hand side of Eq. \ref{Eq:Dual_Fermion_O3_Z2} describes a dual theory with a free Majorana fermion plus a decoupled $Z_2$ gauge theory. In fact, this lattice duality can be viewed as the one studied in Sec. \ref{Sec:ExactMap} with the $Z_2$ global symmetry (introduced in Sec. \ref{Sec:Global_Sym}) promoted to a dynamical $Z_2$ gauge theory.

When we tune $M'=3$ and keep the coupling constant $t$ in Eq. \ref{Eq:Z2_Gauge_Action} of the $Z_2$ gauge field sufficiently large (and positive), the dual theory is at a critical point that contains a massless Majorana fermion and a decoupled deconfined $Z_2$ gauge theory in the IR. Now, we turn to the discussion of the IR nature of the model Eq. \ref{Eq:Lattice_Boson_Fermion_theory_O3_Z2} at this critical point. When $M'=3$, the mass parameter $M$ is always $M<3$. When we choose the coupling constant $J$ such that $1<M<3$, the Chern number $C$ of the bare band structure of the Majorana fermion fields $\chi_n$ in the model Eq. \ref{Eq:Lattice_Boson_Fermion_theory_O3_Z2} becomes $C= 1$. In this regime, if we integrate out the Majorana fermions $\chi_n$ in Eq. \ref{Eq:Lattice_Boson_Fermion_theory_O3_Z2} while neglecting the effect of $S_{\rm int}'$ in this integration, an $\O(3)_{1,1}^0$ Chern-Simons term will be generated and the resulting theory will be naturally identified as a vector boson coupled to an $\O(3)_{1,1}^0$ Chern-Simons gauge theory. However, it is unclear how exactly the interaction terms $S_{\rm int}'$, which is inevitable by the condition Eq. \ref{Eq:Cancellation_Condition}, affects this statement. An observation is that, in a naively continuum limit, all the terms in $S_{\rm int}'$ contain high powers of spacetime derivatives and may be considered irrelevant in a continuum field theory. Also, we notice that the $\O(3)_{1,1}^0$ Chern-Simons term does not have any continuous tuning parameters. These two observations make it plausible that the $\O(3)_{1,1}^0$ Chern-Simons term is not affected by the interactions $S_{\rm int}'$ when we integrate out the fermion $\chi_n$ with a bare mass $3-M$ of order 1 (and with $M-1$ order 1 as well) in the model Eq. \ref{Eq:Lattice_Boson_Fermion_theory_O3_Z2}. Therefore, we speculate that the theory Eq. \ref{Eq:Lattice_Boson_Fermion_theory_O3_Z2} still corresponds to the theory of a vector boson coupled to an $\O(3)_{1,1}^0$ Chern-Simons gauge theory in the IR despite of the interactions $S_{\rm int}'$. If this speculation is correct, the lattice duality discussed in this appendix can be viewed as an UV regulated version of the IR duality between a critical vector boson coupled to an $\O(3)^0_{1,1}$ Chern-Simons gauge theory and a free Majorana fermion plus a decoupled $Z_2$ gauge field proposed in Ref. \onlinecite{Cordova2017}.

\end{document}